\documentclass[sigconf]{acmart}

\AtBeginDocument{%
  \providecommand\BibTeX{{%
    \normalfont B\kern-0.5em{\scshape i\kern-0.25em b}\kern-0.8em\TeX}}}

\copyrightyear{2024}
\acmYear{2024}
\setcopyright{acmlicensed}\acmConference[UIST '24]{The 37th Annual ACM Symposium on User Interface Software and Technology}{October 13--16, 2024}{Pittsburgh, PA, USA}
\acmBooktitle{The 37th Annual ACM Symposium on User Interface Software and Technology (UIST '24), October 13--16, 2024, Pittsburgh, PA, USA}
\acmDOI{10.1145/3654777.3676390}
\acmISBN{979-8-4007-0628-8/24/10}

\newcommand{\haoxiang}{\textcolor{black}}
\newcommand{\xingbo}{\textcolor{black}}
\newcommand{\peng}{\textcolor{black}}
\newcommand{\pzh}{\textcolor{black}}

\newcommand{\zhenhui}{\textcolor{black}}
\newcommand{\fhxr}{\textcolor{black}}

\usepackage{soul}  
\usepackage{color} 
\usepackage{xspace}
\usepackage{listings}
\usepackage{multirow}
\usepackage{multicol}
\usepackage{float}

\usepackage{appendix}  

\newcommand{\eg}{{\it e.g.,\ }}

\newcommand{\ie}{{\it i.e.,\ }}

\definecolor{activegold}{RGB}{255,193,61}

\definecolor{lightorange}{RGB}{230, 170, 50}
\definecolor{lightgreen}{RGB}{121,210,121}
\definecolor{lightteal}{RGB}{121,199,210}
\definecolor{lightblue}{RGB}{100,212,239}
\definecolor{lightpurple}{RGB}{153,102,255}
\definecolor{lightred}{RGB}{245, 132, 120}
\definecolor{red}{RGB}{178,34,34}
\definecolor{gray}{RGB}{166,166,166}

\newcommand{\name}[1]{\textit{LessonPlanner}}

\begin{document}


\title{\pzh{\name{}: Assisting Novice Teachers to Prepare Pedagogy-Driven Lesson Plans with Large Language Models}}
\author{Haoxiang Fan}
\email{fanhx6@mail2.sysu.edu.cn}
\orcid{0009-0000-5729-8491}
\affiliation{%
  \institution{Sun Yat-sen University}
  \city{Guangzhou}
  \country{China}
}

\author{Guanzheng Chen}
\email{chengzh59@mail2.sysu.edu.cn}
\orcid{0000-0002-0152-9120}
\affiliation{%
  \institution{Sun Yat-sen University}
  \city{Guangzhou}
  \country{China}
}

\author{Xingbo Wang}
\email{xiw4011@med.cornell.edu}
\orcid{0000-0001-5693-1128}
\affiliation{%
  \institution{Cornell University}
  \city{New York}
  \state{NY}
  \country{United States}
}

\author{Zhenhui Peng}
\authornote{Corresponding author.}
\email{pengzhh29@mail.sysu.edu.cn}
\orcid{0000-0002-5700-3136}
\affiliation{%
  \institution{Sun Yat-sen University}
  \city{Guangzhou}
  \country{China}
}

\renewcommand{\shortauthors}{Haoxiang Fan and Guanzheng Chen et al.}

\begin{abstract}
\pzh{
  Preparing a lesson plan, 
  \eg a detailed road map with strategies and materials for instructing a 90-minute class, 
  is beneficial yet challenging for novice teachers. 
  Large language models (LLMs) can ease this process by generating adaptive content for lesson plans, which would otherwise require teachers to create from scratch or search existing resources.
  In this work, we first conduct a formative study with six novice teachers to understand their needs for support of preparing lesson plans with LLMs. 
  Then, we develop \name{} that assists users to interactively construct lesson plans with adaptive LLM-generated content based on Gagne's nine events. 
  Our within-subjects study ($N=12$) shows that compared to the baseline ChatGPT interface, \name{} can significantly improve the quality of outcome lesson plans and ease users' workload in the preparation process. 
  Our expert interviews ($N=6$) further demonstrate \name{}'s usefulness in suggesting effective teaching strategies and meaningful educational resources. 
  We discuss concerns on and design considerations for supporting teaching activities with LLMs. 
}
\end{abstract}

\ccsdesc[500]{Human-centered computing~User interface toolkits}
\ccsdesc[500]{Computing methodologies~Natural language generation}
\ccsdesc[300]{Applied computing~Computer-assisted instruction}
\keywords{\fhxr{Large language models, lesson plan preparation, pedagogy-driven system}}



\maketitle
\section{Introduction}
\begin{figure}
    \centering
    \includegraphics[width=0.9\linewidth]{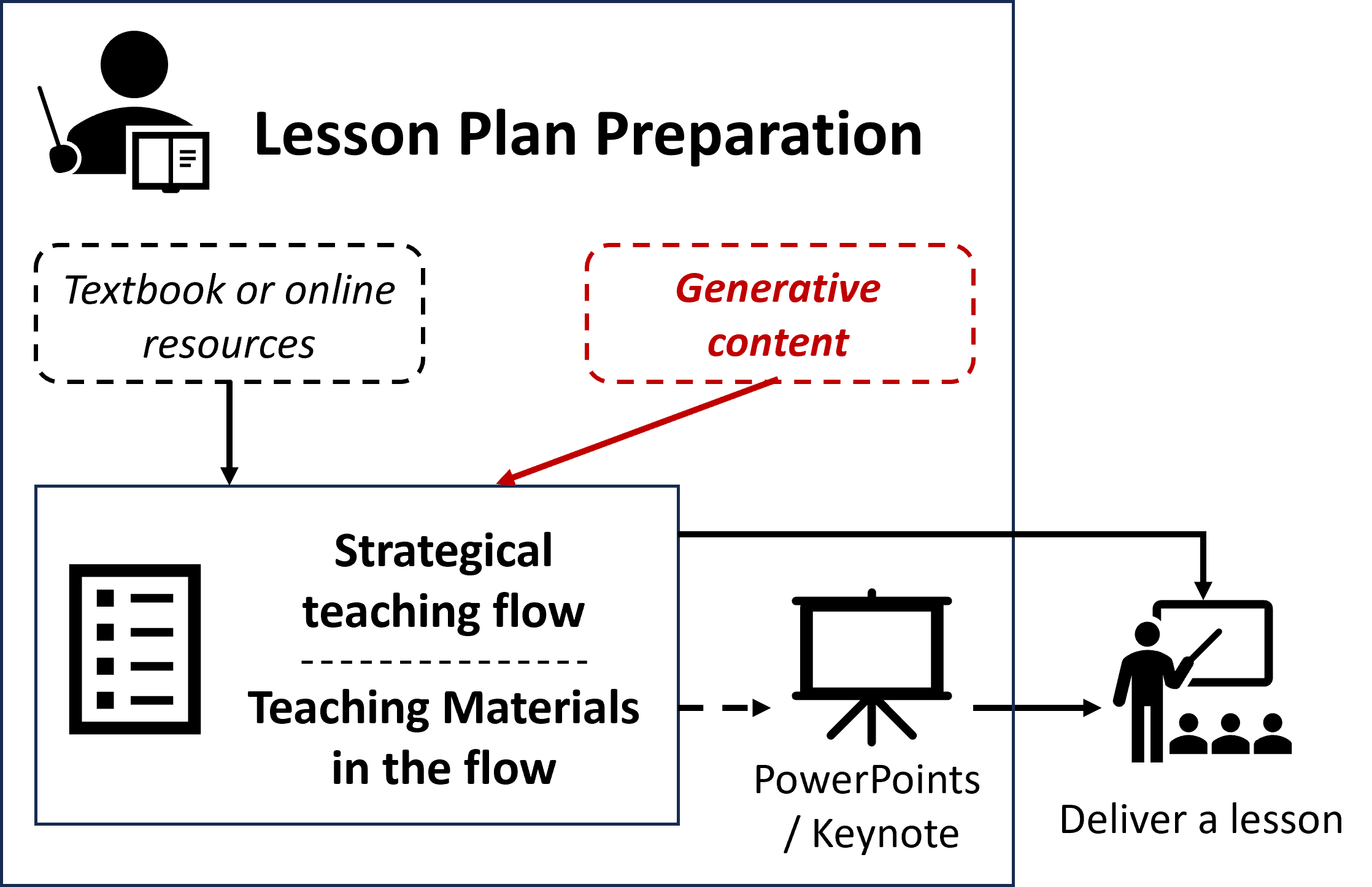}
    \caption{An example lesson plan preparation process. We build an interactive system \name{} to assist teachers with generated content to prepare a documented lesson plan, which can be optionally transformed into slides by users and delivered in their courses.}
    \Description{An example lesson plan preparation process. We build an interactive system \name{} to assist teachers with generated content to prepare a documented lesson plan, which can be optionally transformed into slides by users and delivered in their courses.}
    \label{fig:lesson_plan_process}
\end{figure}

\pzh{
Effective teaching often stems from a well-prepared and organized lesson plan~\cite{wood1988adapting}, which serves as a teacher's detailed description of the course of instruction or learning~\cite{jogan2019effective}. 
\peng{In this paper, we focus on the process of preparing a documented plan that normally describes the teaching flow and materials ~\cite{nesari2014important} for each lesson (\eg 45-minute or 90-minute). 
Such a documented plan can serve as the basis for users to optionally create slides in PowerPoint or Keynote to deliver the lesson,  \haoxiang{or to be directly used as speaking notes in the class} (\autoref{fig:lesson_plan_process}).}
However, constructing an effective lesson plan could be challenging for teachers, especially those who are novices or teaching a course for the first time~\cite{ranawat2021collectiveteach}. 
\fhxr{For one thing, novice teachers lack experience in delivering a course using effective strategies~\cite{ball2007instructional,dias2017challenges}, \eg Gagne's Nine Events of Instruction~\cite{gagne1985conditions}, which helps teachers to prepare and deliver instructional content while considering and addressing the students' conditions, like gaining attention and presenting a stimulus at the beginning of a class.}
For another, even when novice teachers have a mind of planning an event or activity (\eg in-class exercises) at a certain time of a course, they could face difficulties in curating relevant materials to execute the activity~\cite{dias2017challenges,stein2020integration,sawyer2018seeking}. 
As a common practice, novice teachers often refer to others' lesson plans and search related materials in textbooks or online~\cite{sawyer2018seeking}. 
Nevertheless, these traditional approaches could be limited by the diversity 
of referred materials and the ability to support teachers in adapting the content in the lesson plan based on their personal thoughts. 
}

\pzh{
Recent advances in large language models (LLMs) show great potential to address these challenges in composing a lesson plan by suggesting in-situ-generated content that adapts to teachers' thoughts. 
In fact, many teachers have explored the usage of LLMs to prepare for their lectures~\cite{mollick2023using,mogavi2024chatgpt}. 
For example, \citet{mollick2023using} suggested the prompts to LLMs to implement effective teaching strategies in the classroom, such as providing diverse examples to help students comprehend abstract concepts and collecting test questions to help students assess their knowledge. 
\haoxiang{However, querying the LLMs in the wild, \eg via the web app of ChatGPT, could lead to a fragmented lesson plan because of the difficulties in organizing all of the LLM outputs.}
\xingbo{
More importantly, due to the lack of experience in teaching and guidance for prompting, the generated content may not be comprehensive and readily usable, thereby requiring iterative refinement~\cite{koraishi2023teaching, Bill2021Artificial,alam2023intelligence}.}
There is a need to build an interactive system with customized features to ease the process of adopting LLMs to prepare a comprehensive lesson plan. 
Nevertheless, little is known about what are the design requirements for such a lesson planning system, how LLMs can help to satisfy these requirements, and how would teachers perceive and collaborate with the system.  
}

\pzh{
In this paper, we introduce \name{}\footnote{\url{https://github.com/fanhaoxiang1/LessonPlanner}}, an interactive system that offers pedagogy-driven generated content to assist users in constructing lesson plans. 
We first conduct interviews with six participants, including three novice teachers and three teaching assistants in the universities, to understand their challenges and needs for support when using LLMs to prepare lesson plans. 
The findings reveal users' demands for the generated content that aligns with effective teaching strategies and flexible interactions with LLMs. 
Based on the derived design goals from the interviews, we build \name{} as a web app powered by the LLM GPT-4.0. 
In \name{}, users can first input basic course information (\eg course name, topic of one lecture) to generate an outline of the lesson plan with teaching strategies 
suggested by Gagne's nine events~\cite{gagne1985conditions}. 
\haoxiang{Then, for each section of the outline, users can extend it with user-specified LLM-generated activities belonging to each instructional event to get suggested teaching materials, and they have the option to customize the materials via prompts. 
At any time of the lesson plan preparation process with \name{}, users can select any content to ask the LLM to explain it and freely query the LLM about anything, including knowledge delivery strategies and presentation suggestions for creating slides. }
}


\pzh{
We first evaluated \name{} via a within-subjects study in which twelve graduate students or senior undergraduates act as teachers and prepare lesson plans for their familiar course topics. 
The results showed that 
\name{} can significantly improve the quality of outcome lesson plans and ease users’ workload in the preparation process.
\haoxiang{
We further conducted expert interviews with five novice teachers (who have less than three years of teaching experience) and one experienced teacher across various educational stages and subjects. 
Experts further demonstrated that \name{} is useful in lesson planning because it offers a well-organized outline and inspiring content. 
We discuss the concerns and implications of our study on facilitating users in teaching activities with LLMs. 
}
}

In summary, this paper has three main contributions. 
\pzh{
First, we present \name{}, an interactive system that leverages generated content to assist teachers in preparing lesson plans. 
Second, via a within-subjects study and an interview study, we offer empirical evidence on \name{}'s effectiveness and usefulness in helping novice teachers prepare lesson plans. 
Third, based on our findings, we offer design implications for future systems that use large language models to assist teachers. 
}





\section{Related Work}
\subsection{Preparation of Lesson Plans}

\pzh{
A lesson plan is the instructor's road map of what students need to learn and how it will be done effectively during class time~\cite{nesari2014important}. 
\peng{
A carefully constructed lesson plan allows teachers to enter the classroom with more confidence and achieve effective teaching outcomes of their courses~\cite{jacobs2008science,hoover1970learning}.
As suggested by the CIPP (Context, Input, Process, and Product) evaluation model~\cite{stufflebeam2000cipp}, an effective lesson should be needs-oriented, resource-adequate, systematically executed, and outcome-focused.}
A carefully constructed lesson plan can reveal the first two aspects of the CIPP model~\cite{aziz2018implementation}, that is, goals aligned with subject and societal demands as Context and a well-organized content plan as Input.
}

\pzh{
However, teachers, especially those who are novices or teach a course for the first time, could find it difficult to construct a well-designed lesson plan.
For example, they may need to spend a lot of time writing a lesson plan from scratch~\cite{dias2017challenges}, finding high-quality resources related to the lesson~\cite{sawyer2018seeking, ball2007instructional}, and thinking of the strategies to teach each knowledge concept~\cite{dias2017challenges}. 
Researchers have explored various methods that can assist in the preparation of lesson plans. 
They have tried using knowledge graphs ~\cite{saad2018requirement, chan2022intelligent, ranawat2021collectiveteach} and recommendation systems~\cite{chau2017content} to integrate existing web resources, which could make it convenient for teachers to find needed teaching materials. 
For example, CollectiveTeach~\cite{ranawat2021collectiveteach} retrieves and collects documents related to a specific topic based on a particular optimization objective, and then rearranges the documents into a coherent lesson plan.
They have also proposed a variety of lesson planning systems~\cite{chau2017content, andre2011ilessonplan, strickroth2019platon, calandra2007electronic,pender2022ai, zain2017collaborative}.
For instance, ~\citet{pender2022ai} have developed the CLEVER digital web platform, which includes Didactic Guidance, Content Management, Platform Services, and the Data section. It assists users in selecting existing content from the platform's library or in creating new content.
~\citet{zain2017collaborative} also proposed CIDS (Collaborative Instructional Design System), a lesson plan design system aimed at assisting teachers in designing and implementing the requirements of 21st-century learning, integrating features of the ASIE (Analyze, Strategize, Implement, and Evaluate) Instructional Design Model and the Professional Learning Community.
Nevertheless, \peng{these previous approaches and systems largely rely on existing high-quality teaching resources related to the lessons, while such resources may not always be available for any lesson that the teachers want to deliver}. 
}

\pzh{
Our work is motivated by the benefits of preparing a carefully constructed lesson plan and is aligned with previous work that aims at easing this preparation process. 
Different from previous work, we explore the usage of machine-generated content to assist lesson plan preparation. 
}

\subsection{Large Language Models in Education}

Researchers have started to explore the usage of large language models (LLMs) in educational settings due to their ability to provide personalized, efficient, and engaging learning and teaching experiences~\cite{mogavi2023exploring}. 
For learners, depending on a student’s needs and learning style, LLMs are able to create self-study and self-assessment materials~\cite{rahman2023chatgpt, chen2024retassist}, such as knowledge flashcards or course-specific questions~\cite{gabajiwala2022quiz, bhat2022towards, biri2023assessing, divito2024tools}. 
For example, since ChatGPT demonstrated strong insights in explaining medical questions without any additional training \cite{kung2023performance}, \citet{divito2024tools} have utilized it to support medical learners in problem-based learning.
Furthermore, LLMs have the capability to encourage students to think critically~\cite{abdelghani2022gpt}
 \eg by
creating chatting chances during learning and adding suitable problems if needed~\cite{denny2024generative}. 

For teachers, LLMs can help to automate the knowledge assessment process and provide personalized feedback to students~\cite{bernius2022machine, persua2022}.  
Additionally, teachers can get support from generated teaching materials including varied examples, instruction notes~\cite{bonner2023large}, explanations~\cite{mollick2023using}, post-class exercises~\cite{shen2021generate}, a list of frequently asked questions~\cite{mitra2024retllm}, and so on. 
For instance, \citet{mitra2024retllm} proposed a system named RetLLM-E that assists educators in acquiring frequently asked questions from students related to their courses and in generating responses.
The system initially retrieves context from student questions previously answered by teachers on a forum and from related course materials. 
It then uses this information to prompt LLMs to generate specific, high-quality, and precise answers to students' questions, which generally have a better quality than other available answers on the forum.

Despite the promising potential, few works have explored using LLMs to help teachers prepare their lesson plans. 
In this paper, we will first work with six novice teachers to understand their opinions about using LLMs for preparing lesson plans and identify their needs for support. 

\subsection{Interactive Systems that Supports Users with Large Language Models}

Human-Computer Interaction (HCI) researchers have proposed a bunch of interactive systems that leverage LLMs to
improve the efficiency and user experience in various tasks, such as group decision making~\cite{chiang2024enhancing, han2023redbot}, software engineering~\cite{liu2023wants, peng2023impact}, human-UI interaction design~\cite{wang2023enabling}, and so on.
Specifically, many HCI researchers have embedded LLMs in their systems to support co-writing tasks~\cite{mirowski2023co, jakesch2023co, petridis2023anglekindling, jiang2023graphologue, lee2022coauthor}, which is similar to our setting of co-writing a lesson plan with LLMs.
For example, 
~\citet{lee2022coauthor} developed \textit{CoAuthor} and conducted a user study, which highlighted that co-writing with LLMs can aid users by enhancing fluency, pooling ideas, and improving writing quality.
\textit{Wordcraft}~\cite{yuan2022wordcraft} consists of an editor and some buttons used to call an LLM to generate various kinds of content. Compared with using a chatting interface, users find out this tool is more helpful and collaborative~\cite{yuan2022wordcraft}. This finding inspires our design by replacing the chat box with simple pre-set buttons. 
Furthermore, \textit{Jamplate}~\cite{yin2024jamplate} is an idea reflection system that integrates LLM responses into the templates originated from the traditional reflection theory. 
It is shown that the well-organized generated content helps users a lot in expanding and refining their ideas~\cite{yin2024jamplate}. 
Just like Jamplate does, we integrate education theory into our system to structure LLM responses \pzh{in a lesson plan template.} 

We contribute an interactive system \name{} that is in line with these previous interactive systems and offer insights into how LLMs can facilitate lesson plan preparation. 
\section{Formative Study}
This study aims to help novice teachers design lesson plans in an efficient way and get high-quality outcomes.
To achieve this, we conduct a formative study with six novice teachers or teaching assistants.
The insights from the study will inform the design goals for \name{}.
The participants, with a mean age of 29.83 ($SD=5.15$), include three females and three males, and we note as P1 to P6. Half of them (P2, P5, P6) are novice teachers with less than three years of teaching experience, and the other half are teaching assistants (P1, P2, P4) who have taught undergraduate students during one to three academic terms. Four of the participants teach courses related to computer science, while P1 teaches Architecture-related courses and P5 teaches Korean Intensive Reading.

\fhxr{
\zhenhui{
\subsection{Pedagogies for Lesson Planning} \label{intro_theory}
}
Before the interviews, we prepared three educational theories usually used to guide teachers in preparing structured lesson plans, which have proven to be effective~\cite{putra2022developing, razzouk2008analysis, abdulwahed2009applying}.
They are easy to implement in the system to assist teachers in designing lesson plans.
\textbf{Bloom's Taxonomy} is a hierarchical classification of cognitive skills that educators use to structure curriculum learning objectives, assessments, and activities, ranging from lower-order thinking skills like remembering, \zhenhui{understanding, and applying,} to higher-order skills like analyzing, evaluating, and creating.
\textbf{Kolb's Learning Cycle} is a four-stage model of experiential learning that emphasizes the process of learning through experience, consisting of concrete experience, reflective observation, abstract conceptualization, and active experimentation.
\textbf{Gagne's Nine Events of Instruction} involve the actions of both teachers and learners throughout the teaching process~\cite{khadjooi2011use}.
The nine events include: 
\zhenhui{
\textbf{Gaining Attention}: Present introductory activities that engage learners;
\textbf{Informing Learners of Objectives}: Clearly state the learning goals and outcomes;
\textbf{Stimulating Recall of Prior Learning}: Encourage learners to remember and connect previous knowledge;
\textbf{Presenting Stimulus}: Introduce new content and information;
\textbf{Providing Learner Guidance}: Offer instructions and strategies to help learners understand and process the new content;
\textbf{Eliciting Performance}: Have learners practice what they have learned;
\textbf{Providing Feedback}: Give constructive feedback on learners' performance;
\textbf{Assessing Performance}: Evaluate learners' understanding and skills;
\textbf{Enhancing Retention and Transfer}: Use activities that help learners retain information and apply it to new situations.
}
The nine events and corresponding activities \zhenhui{prepared for the discussion in the formative study} are summarized in~\autoref{tab:gagnes}.
Each event may occur sequentially in a class as described above, but they can also be repeated to better organize the instructional content.}


\subsection{Procedure}
To assist novice teachers in solving the difficulties in lesson planning and offer a better user experience, we conduct a two-phase formative study to comprehensively understand the users' needs. 
Initially, the participants are involved in semi-structured interviews.
A set of questions is presented, and they are encouraged to express their viewpoints. 
The discussions are intended to investigate the processes involved in creating their ways of routine lesson plan creation (\eg ``Do you typically design instructional notes or slides for a 90-minute class?'', ``Where do you like to access related materials when developing lesson plans?''), the potential of integrating \zhenhui{the three educational theories (introduced in~\autoref{intro_theory})} into a lesson plan (\eg ``Do you agree that Gagne's Nine Events can effectively guide instruction?''), and the challenges encountered during the design process (\eg ``What challenges do you face during this process?'').
Furthermore, we specifically investigate their perspectives regarding LLMs such as ChatGPT in the design of lesson plans. Each interview has a duration of around 45 minutes in this phase.

Subsequently, they are invited to participate in a co-design session aimed at exploring and evaluating various interface designs.
The objective is to identify the user interface requirements for a system designed for lesson plan design. Four slide decks (shown in Appendix A) are presented to participants, featuring example designs. These include a page for input meta-data of class, an outline overview page, and two editing pages with LLMs in different forms. Participants are provided with the ability to resize and crop screenshots, draw shapes, and use text boxes for the purpose of designing.
One author assists with sketching by making edits based on participants' responses. We also ask questions to provide more detailed explanations of their ideas or explore different design options. Each participant spends about 15 minutes in this session.


\pzh{\subsection{Findings}}
\pzh{
We use the reflexive thematic analysis method \cite{braun2012thematic} to analyze the transcribed recordings of each participant's semi-structured interviews and co-design sessions. 
\peng All mentioned that they primarily prepared a lesson plan for a lecture in the form of PowerPoint slides. 
\peng P2 said that he sometimes chose to annotate some text in the slides or textbook to remind themselves of extra information, such as proof of formulas or supplemental examples. Except for P2, no participants reported that they left notes under the slides or in a separate paper as a reminder of critical information during the lecture, such as the logic of proof and supplemental examples.  
\peng{Nevertheless, they all agreed that having a word-like document to list the main flow and content of a lecture was helpful, as they could \textit{``easily build up slides based on the documented lesson plan''} (P6)}. 
We summarize the participants' challenges in lesson plan preparation and using large language models (LLMs) to assist the preparation as below.
}



\pzh{
\subsubsection{Challenges in Lesson Plan Preparation}
\pzh{
\textbf{C1. Lack of adaptive support in planning effective teaching strategies in the lessons.} 
All participants said that they mostly relied on their teaching experience and others' plans to design their teaching strategies in a lesson. 
Nevertheless, after we introduced Gagne's Nine Events of Instruction (the list of the events is shown in~\autoref{tab:gagnes}), all participants agreed that they actually adopted some of the events in their lectures. 
\textit{``I did not get official training on educational or teaching theories, but I found that I had enacted several strategies in the suggested events of instruction. These nine events will be generally helpful in my course''} (P2). 
\fhxr{
Compared to the other two theories mentioned in~\autoref{intro_theory}, two participants (\ie P1, P2) confirmed that Gagne's theory was more feasible to be embedded in the system because \textit{``it is relatively more specific and provides teachers with comprehensive guidelines''} (P2). Furthermore,
}
four participants (\ie P1, P3, P4, and P6) with insufficient teaching experience further mentioned that they would like to exercise all the nine events in their lectures but they lack adaptive support in incorporating these events into a specific lesson.
\textit{``I would like to try the suggested `gaining attention' and `providing learner guidance', but the example usages of these strategies could not be directly adopted in my \peng{Experiments of Data structures course''}} 
\peng{(P1)}. 
}
}

\pzh{
\textbf{C2. Time-consuming process in searching for adequate teaching materials.} 
All participants indicated that preparing for a lesson, especially the one they taught for the first time, is a time-consuming process, where they spent most of the time preparing the teaching materials. 
\textit{``I usually need to spend a considerable amount of effort finding and determining the proper materials that support the teaching activities in my course, \eg finding a good case to illustrate the real-world application of the taught concepts''} (P5). 
Even though they can get started from textbooks or others' lesson plans, all participants expressed their desire to customize the teaching flow and materials based on their thoughts. 
As such, they spent time looking for related high-quality materials online or creating needed materials by themselves. 
\textit{``I was used to prepare an in-class exercise after explaining a key concept. I normally search for suitable exercises online but the returned results were often disorganized or unrelated. I had to pay a lot of attention to identify the ones I need''} (P1).
}



\pzh{
\subsubsection{Challenges on Using LLMs in Lesson Plan Preparation}
\textbf{C3. High mental demand in manipulating prompts to get related content.} 
\peng Three participants (P3, P4, and P6) had experiences using LLMs (Specifically, ChatGPT) to prepare for their lessons. 
For example, \peng{P3} reported that \peng{she} had used ChatGPT to \peng{generate precise answers to the questions that students may ask when delivering knowledge points}. 
The others also see the potential of using LLMs to generate suggestions on the teaching flow and materials based on their needs.  
However, P5 and P6 commented that it was or would be mentally demanding to think of the prompts for LLMs to get needed content. 
\peng{\textit{``Searching for information directly seems more efficient than querying ChatGPT with specific questions.''} (P6). \textit{`` If LLMs cannot provide satisfying materials for me instantly, I prefer not to engage in back-and-forth dialogue to refine the results''}} (P5).
}



\textbf{C4. Distrust of the usefulness of generated content}. 
Except for P3, the remaining five participants conveyed a sense of distrust toward the outputs of the LLM. On one hand, some (P2, P5, P6) doubted the accuracy of the content. \textit{``I don't trust the detailed content generated by LLMs, and I would definitely double-check it before using it. Therefore, I generally only use LLMs to create a rough outline, and I search for the specific content myself''} (P1).
On the other hand. some (P1, P2, P6) have raised concerns about the expertise of LLMs in specific subjects, suggesting that they may struggle to generate useful content.\textit{``ChatGPT can generate many correct questions, but that doesn't mean they are good questions in this course (Computer Graphics). Because it may lack a deep understanding of the subject''} (P6).




\subsection{Users' Preference on the Interface}
During the co-design phase, all participants reached a consensus that the layout of our outline overview page, which organizes the outline into distinct blocks based on the generated subtitles, was a good design. P4 advised that \textit{``the instructional events are important cues for me, and they should be more eye-catching.''}. 
For the course metadata collection page, P2 questioned that \textit{``the available data input was excessively restricted. I believe it is necessary to enter the course designed for graduate or undergraduate students. Without this information, how can the system accurately determine the specific type of content to generate?''} Moreover, participants were requested to compare the LLM assistant interface integrated within the editor as opposed to a sidebar interface. P1, P2, P3, P5, and P6 expressed a preference for the LLM as a sidebar. \textit{``It resembled common design practices''} (P3).
\subsection{Teachers' Needs for Generated Content in Lesson Plans}
\peng{
\autoref{tab:gagnes} summarizes the particular instructional activities our participants commonly incorporate during each of Gagne's Nine Events in the classroom.}
Additionally, we ask their perspectives on the capacity of LLMs to contribute to the development of materials for the activities they have mentioned in each interview. 
The results are displayed in the third column of \autoref{tab:gagnes}, with `Y' if they think the LLMs can facilitate the activity and `N' if they can not.
The results indicate that, with the exception of subject-specific activities (e.g., providing source code, providing example sentences), most teachers organize comparable activities for a particular instructional event.
For the majority of classrooms, LLM can help teachers in creating relevant instructional materials. This highlights the potential of LLMs to serve as assistants in lesson planning.

\begin{table*}[]
\caption{Nine events of Gagne's instructional theory and the activities that our participants in the formative study suggest to use in each event. 
Participants also share their opinions on whether (Yes or No) LLMs are able to help them in each activity. 
In \name{}, we have preset template prompts to LLMs for most of the suggested activities. 
}
\label{tab:gagnes}
\begin{tabular}{cccc}
\hline
\textbf{Gagne's Event}                              & \textbf{Common activities or resources in this event}  & \textbf{LLMs facilitated} & \textbf{Implemented} \\ \hline
Gain attention                                      & Pose open-ended questions or case studies                & Y                         & Y                    \\ \hline
\multirow{2}{*}{Inform learners of objectives}      & Create ordered lists of knowledge points                 & Y                         & Y                    \\
                                                    & Display table of contents in slide                       & Y                         & Y                    \\ \hline
\multirow{2}{*}{Stimulate recall of prior learning} & Compile prerequisite knowledge list                      & Y                         & Y                    \\
                                                    & Provide prerequisite knowledge examples                  & Y                         & Y                    \\ \hline
\multirow{5}{*}{Present stimulus}                   & Provide the definition                                       & Y                         & Y                    \\
                                                    & Provide algorithms                                               & Y                         & Y                    \\
                                                    & Provide source code                                              & Y                         & Y                    \\
                                                    & Provide equations                                                & Y                         & Y                    \\
                                                    & Provide a example sentence                             & Y                         & N                    \\ \hline
\multirow{3}{*}{Provide learner guidance}                            &  Explain examples in detailed                                     & Y                         & Y                    \\
                                                    & Design animations in PowerPoint                          & N                         & -                    \\
                                                    & Play videos                                              & N                         & -                    \\ \hline
\multirow{3}{*}{Elicit performance}                 & Construct multiple choice or fill-in-the-blank questions & Y                         & Y                    \\
                                                    & Propose open-ended questions                             & Y                         & Y                    \\
                                                    & Construct group discussion topics                        & Y                         & Y                    \\ \hline
Provide feedback                                    & Offer problem solutions                                  & Y                         & Y                    \\ \hline
Assess performance                                  & Assign homework                                                 & Y                         & Y                    \\ \hline
\multirow{2}{*}{Enhance Retention and Transfer}     & Assign Projects as homework.                                  & Y                         & Y                    \\
                                                    & Select topics for writing papers                                & Y                         & N                    \\ \hline
\end{tabular}
\end{table*}

\pzh{
\subsection{Design Goals of \name{}}
Our participants actively offered suggestions on the design of \name{} to address their challenges. 
Based on their suggestions and related work, we derive the following four design goals (DGs) of \name{}. 
}

  \peng{\textbf{DG1: \name{} should encourage and facilitate teachers to apply effective teaching strategies in the planned lessons.}}
Due to the insufficient adaptive support provided by teachers in lesson planning for the development of effective teaching strategies (C1), \name{} to motivate and assist teachers in applying efficient teaching strategies, especially those who lack experience in teaching.
  
  \peng{\textbf{DG2: \name{} should generate high-quality materials adapted to the planned teaching activities and provide guidance on how to deliver these materials.}}
  The process of searching for suitable teaching materials is time-consuming (C2). Hence, \name{} is expected to produce high-quality materials that are in line with the teaching activities and offer suggestions on how to effectively deliver these materials in the classroom.

  \peng{\textbf{DG3: \name{} should offer pre-set prompts about what users often want to ask the large language models for.}}
  The utilization of LLMs for lesson planning requires a high cognitive load (C3). Therefore, \name{} offers pre-set prompts to assist users in getting materials more efficiently from LLMs.

  \peng{\textbf{DG4: \name{} should provide flexible user control for interacting with the large language models and editing the lesson plans.}}
  Granting users the autonomy to select their preferred method of managing the content generated by LLMs, and to discard, modify, or regenerate it as needed, can mitigate users' distrust towards LLMs (C4).
  

\section{\name{}}
\peng{
In this subsection, 
we present the design and evaluation of \name{}, an interactive system that supports novice teachers to create lesson plans with large language models (LLMs).
\name{} is designed as a web application. 
The front end is implemented using Vue 3 and JavaScript, while the back end is created with Python FastAPI.
Furthermore, we opt to utilize the OpenAI gpt-4-1106-preview as the LLM to provide generated content in \name{}. 
The front end is in Chinese, so the Edge translation plugin was utilized to translate the website in order to present the illustrations in this chapter.
}

\begin{figure*}[tbp]
  \centering
  \includegraphics[width=1\textwidth]{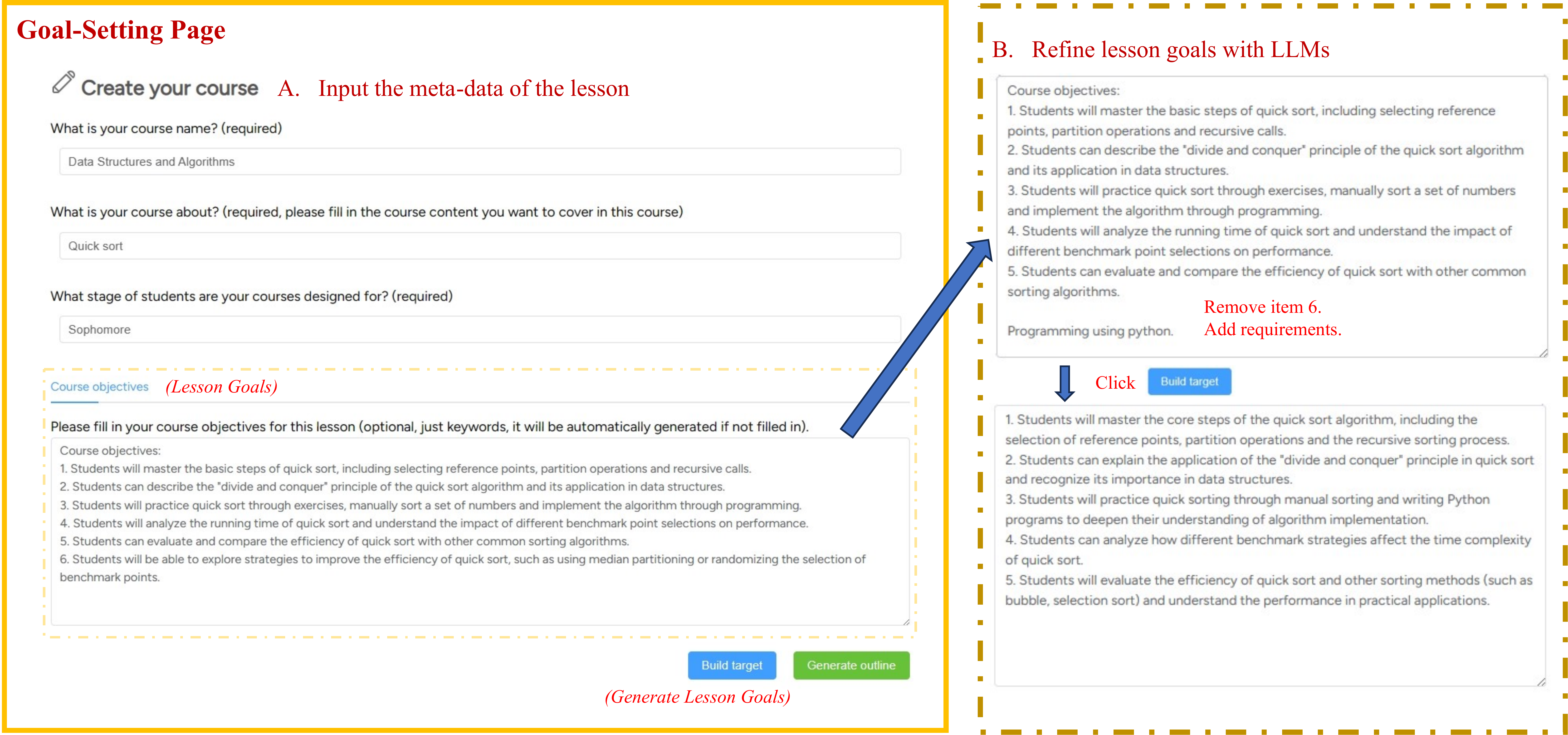}
  \caption{Goal-setting page of \name{}, with the illustration of the process of refining lesson goals with LLMs. The text in parentheses serves as a correction to the mistranslated output.}
  \label{fig:goal}
  \Description{Goal-setting page of \name{}, with the illustration of the process of refining lesson goals with LLMs. The text in parentheses serves as a correction to the mistranslated output.}
\end{figure*}

\begin{figure*}[tbp]
  \centering
  \includegraphics[width=1\textwidth]{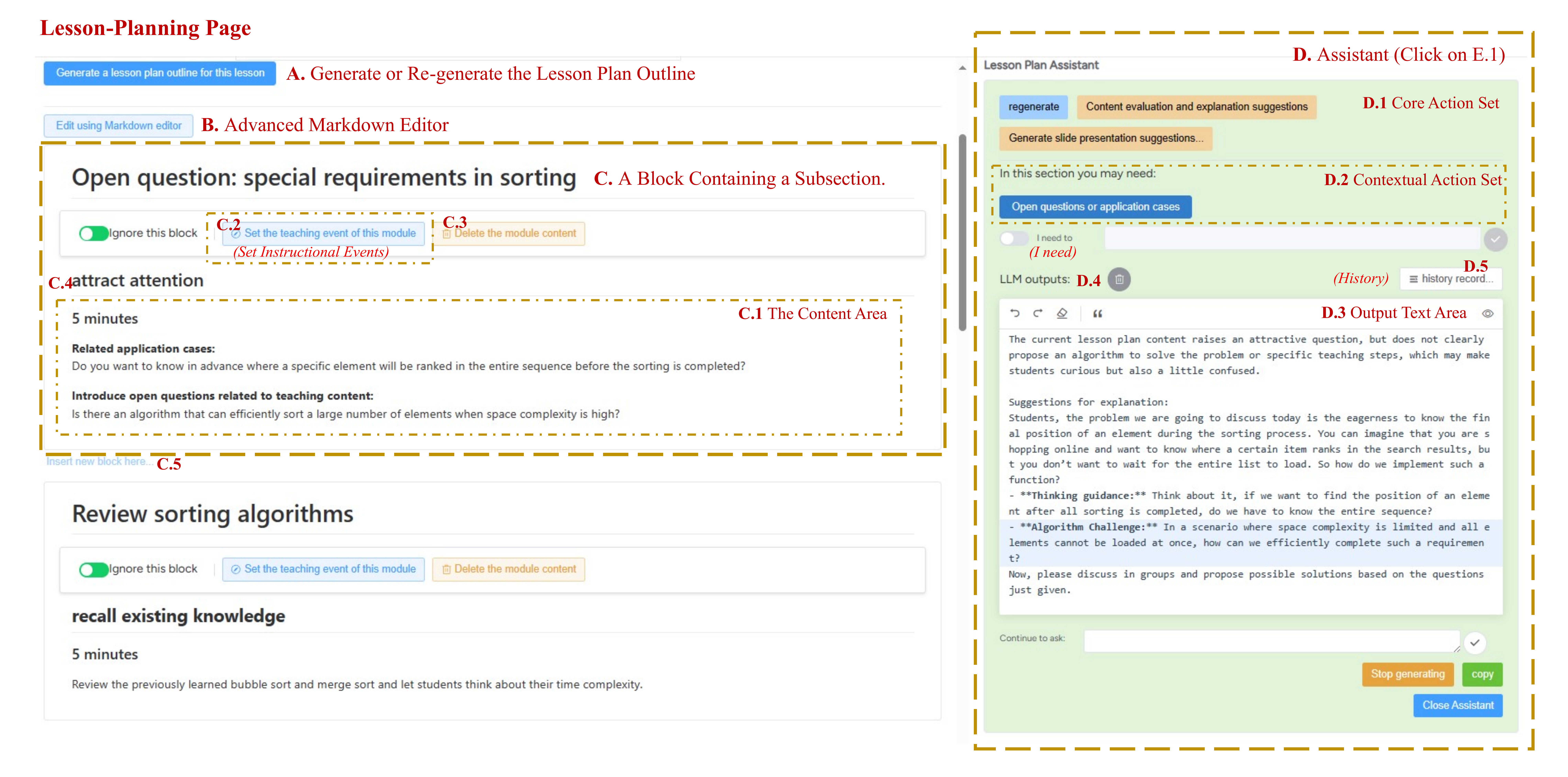}
  \caption{\fhxr{The lesson-planning interface for \name{}. The text in parentheses serves as a correction to the mistranslated output.}}
  \Description{The lesson-planning interface for \name{}. The text in parentheses serves as a correction to the mistranslated output.}
  \label{fig:edit}
\end{figure*}

\begin{figure}[tbp]
  \centering
  \includegraphics[width=0.48\textwidth]{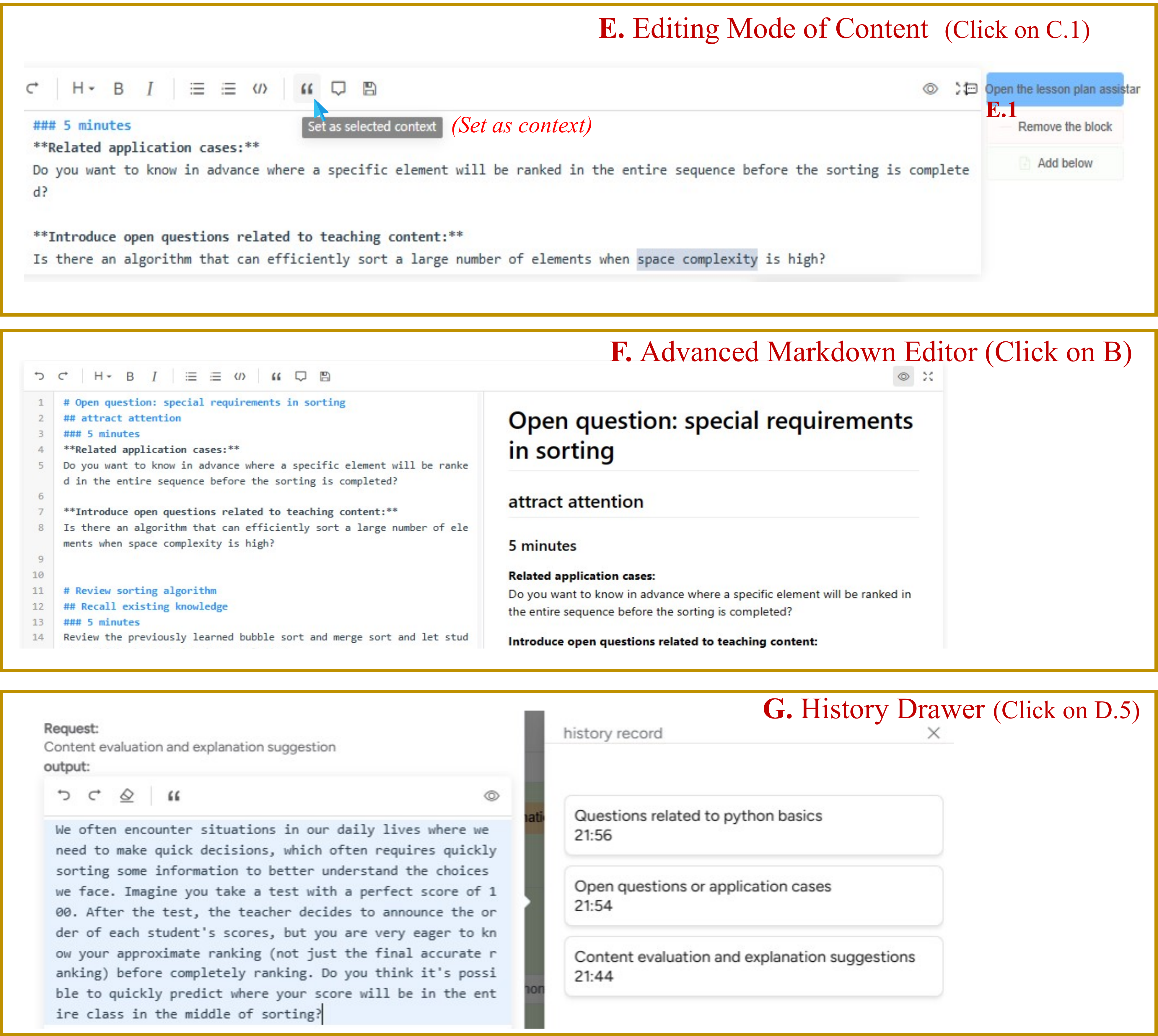}
  \caption{\fhxr{The subsequent changes or pop-up windows triggered by clicking a button on~\autoref{fig:edit}.}}
  \Description{The subsequent changes or pop-up windows triggered by clicking a button on the edit interface.}
  \label{fig:edit_trigger}
\end{figure}

\begin{figure}[tbp]
  \centering
  \includegraphics[width=0.5\textwidth]{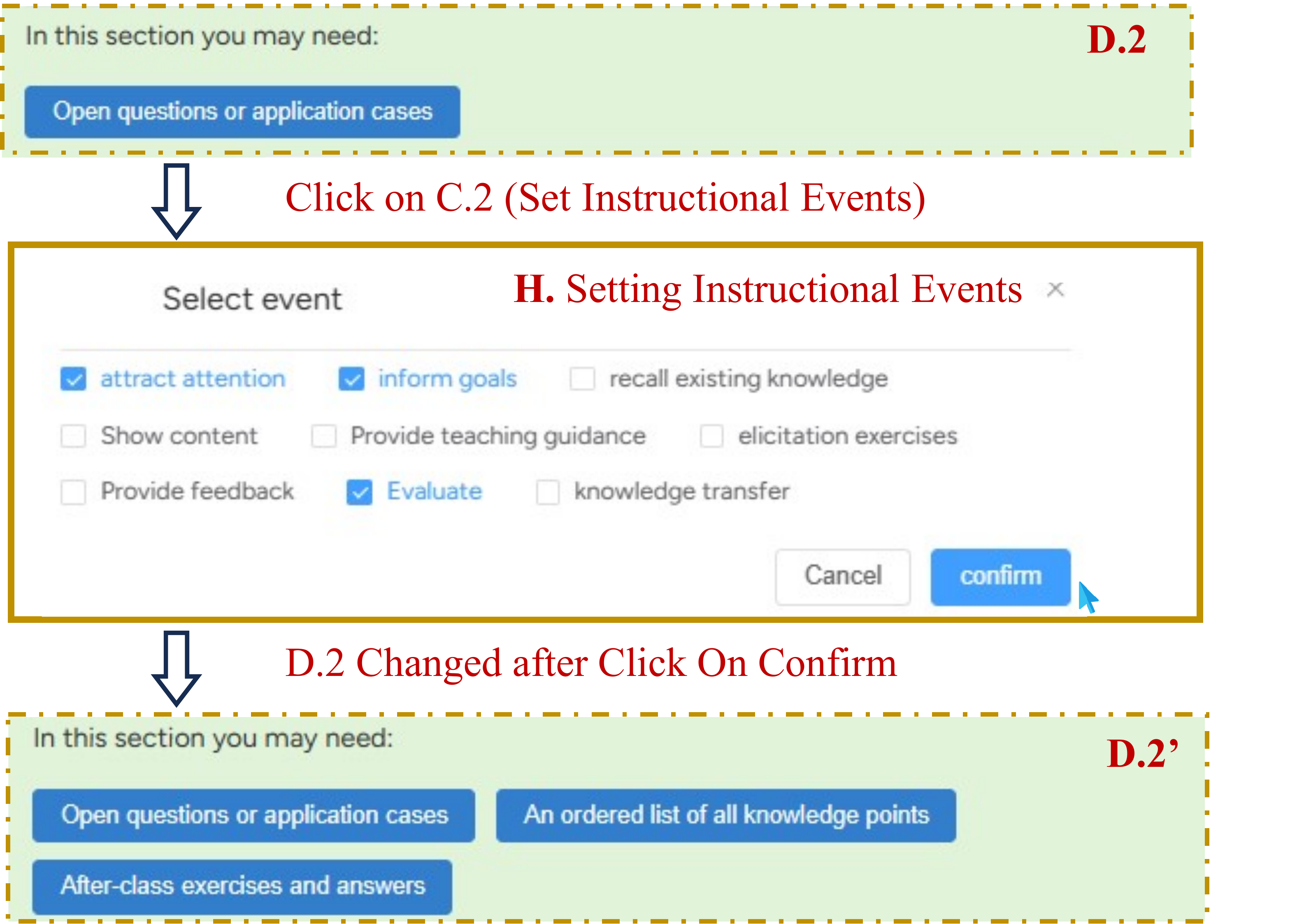}
  \caption{\fhxr{The process of setting instructional events by users in the lesson-planning interface.}}
  \Description{Set instructional events by users in the lesson-planning interface.}
  \label{fig:edit_set_events}
\end{figure}

\peng{
We carefully designed \name{}'s interface and interaction based on the design goals obtained from the formative study. 
To plan a lesson with \name{}, users can engage with the LLM to set teaching objectives to adhere to Bloom's Taxonomy~\cite{forehand2010bloom} and generate an initial outline of the lesson plan conditioned on Gagne's Nine Events (DG1, \pzh{\autoref{fig:goal}}). 
\pzh{
For each section in the outline, users can see suggested teaching materials and strategies to enact this event (DG2, \pzh{\autoref{fig:edit}: C1, D2}). 
Based on the identified teachers' needs for generated content in lesson plans in formative study,  \name{} presets some activities for each event about what users often want to ask the LLM for (DG3, \pzh{D1, D2}). 
Besides, \name{} also offers users flexible options to regenerate the content of an event and copy it to the editor (D), freely edit the content (\autoref{fig:edit_trigger}: E, F), select wanted events (\autoref{fig:edit_set_events}), check the history of generated content (\autoref{fig:edit_trigger}: G), and so on (DG4).
In the following subsections, we will detail the design and implementation of the key features in \name{}. 
}
}

\peng{
\subsection{Goal-Setting Page}
As shown in \autoref{fig:goal}A, on the goal-setting page, users need to input the course name (\eg Data Structures), the topic of the planned lesson (\eg Quick Sort), and the specific stage of the lesson tailored for (\eg Sophomore (2nd-year undergraduate)). 
This meta information is utilized by \name{} in all predefined prompts to facilitate content generation. 
}

\peng{
Users can input their lesson goals, or, in a more convenient way, click the ``Generate Lesson Goals'' button to first check the generated goals conditioned on Bloom's Taxonomy. 
They can then modify the generated goals, \eg remove item 6 and add requirements like ``programming using python'' (\autoref{fig:goal}: B), and click the ``Generate Lesson Goals'' again to iterate the design goals with LLM. 
Once users are satisfied with the lesson goals, they can click on the ``Generate outline'' button, which will initialize a lesson plan outline conditioned on Gagne's Nine Events. 
Users now can proceed to the Lesson-Planning page (\autoref{fig:edit}), as described in the next subsection. 
}

\peng{
\subsection{Lesson-Planning Page}
\subsubsection{Interactive block-based lesson plan outline}
The generated lesson plan outline is displayed in the form of block-based interactive markdown text (\autoref{fig:edit}: C). 
In each block, the editable h1 heading, \eg ``Open question: special requirements in sorting'', is a summary of the content in one section in the planned lesson. 
Each section contains one or more suggested events (C4) that teachers can plan in this section. 
The events are sourced from Gagne's Nine Events and separated by editable h2 headings (\eg ``attract attention'') with an h3 heading (\eg 5 minutes) indicating the planned time for this event.
In each section, there are related teaching materials and strategies, \eg \textbf{Related application cases} ``Do you want to know in advance where a specific element will be ranked in the entire sequence before the sorting is completed?''. 
As a lesson may not necessarily cover all of the suggested sections and may include other sections the users need, users have the options to ignore or specify needed sections, delete the section (C3), insert a new block, and specify the title of a new section at any location (C5). 
Users can also regenerate the lesson plan outline (A), \eg if they are not satisfied with the current one. 
}

\peng{
Users can click the content area (\autoref{fig:edit}: C1) of each block to invoke the editing mode of this section (\autoref{fig:edit_trigger}: E) or click the ``Edit using Markdown editor'' button to invoke an advanced Markdown editor of all content in the lesson plan (\autoref{fig:edit_trigger}: H). 
The editor provides features like ``bold'', ``headings'', and ``lists'' that common editors have. 
Upon finishing the edition of the lesson plan, users can click the ``Download Lesson Plan'' button (not shown in \autoref{fig:edit}) at the bottom of the Lesson Planning page to download the lesson plan as a markdown file. 
}

\peng{
\subsubsection{LLM assistant}
In the markdown editor of each blocked section (\autoref{fig:edit_trigger}: E), users can click the ``Open the lesson plan assistant'' button (\autoref{fig:edit_trigger}: E1) to invoke the LLM assistant in the sidebar (\autoref{fig:edit}: D) that helps users refine the content in this section.  
The assistant includes a Core Action set (D1), Contextual Action set (D2), and Output Area (D3).
The Core Action set contains a collection of preset buttons designed to call functions that regenerate the content in this section, evaluate the content give instructional suggestions, and advice on presenting it, and give suggestions on making a slide based on it. 
The Contextual Action set consists of buttons that list all the preset activities of events (C4) under this section. \autoref{tab:gagnes} displays the implemented activities for each event in our system, derived from the results of the User Study, excluding activities that were difficult to implement because of the needed various prompts in different subjects. 
If clicking on ``Set Instructional Events'' (\autoref{fig:edit}: C2) to change events in the current section (\autoref{fig:edit_set_events}: F), the activities in the Contextual Action set will also change (\autoref{fig:edit_set_events}: D2 change into D2').
If users prefer to freely inquire about LLM, they can enable the ``I need'' button and question the LLM. 
The Output Area contains a Markdown editor (\autoref{fig:edit}: D3) that displays the LLM outputs after users click any button or freely query LLM in the Action sets.
It also has a text area under D3 for users to conduct multiple rounds of interaction with LLM on the content in the markdown editor.  
When interacting with LLM via either preset buttons or text area, users can select any text in the editor and click the ``(set as context)'' button (\autoref{fig:edit_trigger}: E) to use the selected text as context to prompt LLM. 
For example, if there are four key knowledge points in the current section, clicking "Generate Multiple Choice Questions" is highly likely to result in the generated questions that include a range of knowledge points.
If users select the text of one knowledge point as context, the generated questions will be likely to be specific to that knowledge point. 
Users can select any content in the LLM outputs\footnote{If nothing is selected, clicking the "Copy" button will copy all the content in the output text area to the pasteboard.} and click the "Copy" button at the bottom of the assistant panel to copy to the pasteboard and then paste it in any position of the editor of this current section.  
Selecting the Trash icon (\autoref{fig:edit}: D4) will delete all content in the Output Area while clicking the "Stop Generating" button can halt the LLM outputs.  
Clicking the "Close Assistant" button will hide the assistant. 
Additionally, clicking the "history record" button (\autoref{fig:edit}: D5) allows users to access previously generated content (\autoref{fig:edit_trigger}: G).
}

\fhxr{
\zhenhui{
\subsection{Implementation Workflow}
}
\name{} incorporates a workflow that chains all prompts and system functions with flexible user control.
}
\fhxr{
\subsubsection{Metadata and generated objectives as slots in prompt templates}
In the Goal-setting page~\autoref{fig:goal}, teachers are able to enter courses' metadata (\eg course name, course topic), which are used as slots in the prompt templates, as shown below.
\begin{itemize}
    \item \{course name\}: \eg Data Structures and Algorithms
    \item \{lesson topic\}: \eg Quick sort
    \item \{students stage\}: \eg Sophomore
    \item \{lesson goals\}: the content input by the user or LLM in the Lesson Goals area in the Goal-Setting Page
\end{itemize}
When designing Prompts for Core or Context Actions set, we always place this paragraph at the beginning of them:
\begin{quote}
    I \zhenhui{will instruct} the course of \{lesson name\} - \{lesson topic\} for students in \{students stage\}. Here are my lesson goals: \{lesson goals\}.
\end{quote}
This setting helps impose strict restrictions on all generated content.
Sometimes, due to the lack of key information, LLMs are unable to infer accurate information about the course from the content of the current section or selected context, resulting in output that is unrelated to the course.
This also applies to the prompts designed for ``I need'' feature or for having continuous conversations with LLM. Before engaging in a conversation, we always place this prompt at the beginning of the user's first request.
}
\fhxr{
\subsubsection{Leveraging the nine events to ensure control over generated content}
As described in DG1, \name{} should provide effective teaching strategies at each stage (\ie outline generation, interacting with LLMs) rather than merely offering educational resources. 
Additionally, displaying the outline in a block-based format for each section and ensuring that LLMs are aware of the corresponding teaching strategies requires a precisely defined workflow.
}

\fhxr{
\textbf{Combine all the instructional events and metadata when generating outlines.}
As seen in \autoref{fig:edit} C4, recommended instructional events are set once a section of the outline is generated.
The preset prompt for generating the outline is provided in supplementary material 1. Developers have confirmed that GPT-4 possesses background knowledge about Gagne's Nine Events \zhenhui{in their trials on the preset prompt.} 
Therefore, only the Nine Events and corresponding activities in \autoref{tab:gagnes} are included in the prompts of each action. 
LLMs have demonstrated superior reading comprehension and the ability to analyze lengthy texts~\cite{touvron2023llama}.
We have fully leveraged this capability to create outlines that are aligned with the course content and incorporate the Nine Events effectively.
}

\fhxr{
To ensure that the format of the contents generated by the LLMs meets the requirement and minimizes errors, the formatting instructions are set as simple as possible.
We defined that a new section should start with a single ``\#'', and instructional events in every section should be specified with ``\#\#''. 
After generation, a \textit{Formatter} then splits and maps the content into several blocks in the user interface. Also, the whole templates and instructional events in the section can be defined as slots, which will be used in some templates.
\begin{itemize}
    \item \{outline\}: generated outline of the lesson plan.
    \item \{current section\}[events]: the events in current section.
\end{itemize}
}

\fhxr{
\textbf{Emphasize solely the instructional events involved in the current section.}
As for interacting with LLMs within a certain section, if the same context or even section content is assigned to different teaching events, we hope to click on the activities in the Core Actions set or use ``I need'' feature to bring results that match the current teaching events.
For example, when clicking on ``evaluate the content and give instructional suggestions'' in the section with the event ``Stimulate recall of prior learning'', we expect a brief explanation.
In contrast, in the section with ``Provision of Learner Guidance'', we anticipate detailed information to be provided about the knowledge.
We dynamically insert the current section's instructional events and their corresponding definitions into the prompts through programming. 
It helps to remind the LLM to generate more relevant content based on these events. 
An example is shown below. 
\begin{quote}
    The educational theories involved in this section are as follows. Be sure to construct \zhenhui{the lesson plan} around the following events. Events that are not mentioned cannot be covered in this section.\\
    (\{current section\}[events] and their definitions, respectively.)
\end{quote}
\zhenhui{Based on our trials}, it is not recommended to include all instructional events and their corresponding definitions and add the current events after that, \zhenhui{which will make the generated content extensively cover each event and cause information redundancy.} 
}
\fhxr{
\subsubsection{\zhenhui{Chaining} the prompts with examples of input and output.} \label{example_prompt_with_input_and_output}
During the process of iteratively designing prompts, we found that relying on descriptive statements frequently results in unstable output.
\zhenhui{To improve the robustness of the output}, we design an example at the end of \zhenhui{every} prompt to ensure that the system generates results that align with expectations every time.
For example, for the activity ``Generate definition'', the example 
}

\begin{quote}
Input:

Context: Quick Sort

Output:
**Quick Sort * *:

-Definition: Quick Sort is a divide-and-conquer algorithm. It works by selecting a 'pivot' element from the array and partitioning the other elements into two sub-arrays, according to whether they are less than or greater than the pivot. \footnote{https://en.wikipedia.org/wiki/Quicksort}
\end{quote}
\fhxr{
was added to the end of the prompt, forcing LLMs not to generate more detailed explanations to further explain the definition.
}

\section{Evaluation}

\begin{figure*}[tbp]
  \centering
  \includegraphics[width=1\textwidth]{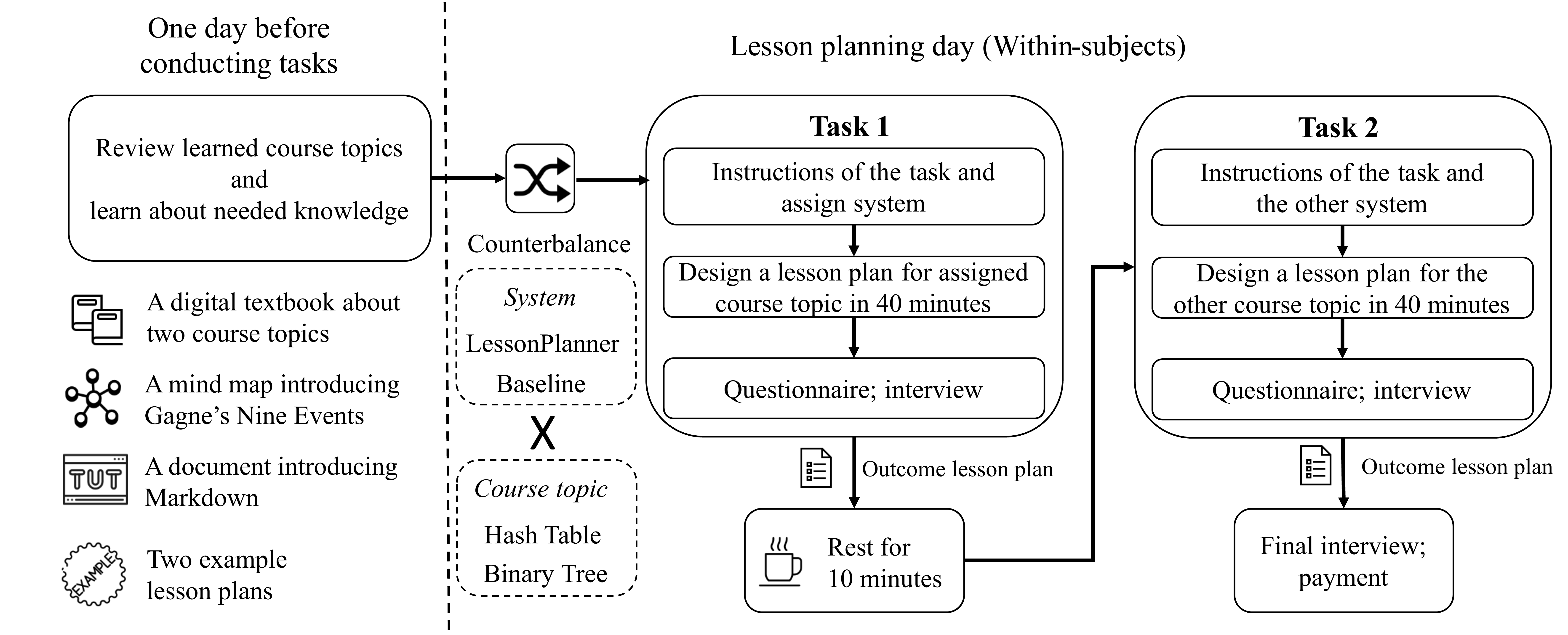}
  \caption{The procedure of the within-subject study.}
  \Description{The procedure of the within-subject study.}
  \label{fig:procedure}
\end{figure*}

\peng{
To explore the effectiveness and user experience of \name{} for assisting novice teachers in lesson planning, we conducted two user studies. 
First, we conduct a within-subjects study with 12 university students who have little or no teaching experience to use \name{} and a baseline tool to prepare lesson plans for a course that they have learned before. 
Second, we conduct an expert interview with 6 teachers from different educational levels and subjects. 
Our research questions (RQs) are:
}

\textbf{RQ1. } How would \name{} affect the lesson planning outcome?

\textbf{RQ2. } How would \name{} affect the lesson planning process? 

\textbf{RQ3. } How would users perceive \name{} for preparing lesson plans?  

\peng{
\subsection{Within-subjects study}
\subsubsection{Participants}
We recruit 12 students (P1-P12, two females, ten males; age: $Mean = 23.42$, $SD=1.98$) via word-of-mouth from a university in mainland China (\autoref{tab:participants}). 
All participants, including two 4th-year undergraduate students and ten graduate students, major in artificial intelligence and are familiar with the course Data Structures and Algorithms. 
The experimental setup controls the lessons that participants need to prepare for, which leads to unbalances of majors and genders of our participants. 
We will discuss the limitations of this setting in the Discussion section. 
Nine of the participants have experience working as a teaching assistant (TA) in the university or as a home tutor for high-school students. 
\pzh{All participants are} quite interested in having a trial on utilizing artificial intelligence tools or online resources for lesson planning ($Mean = 4.33$, \pzh{$SD = 0.47$}; 1 - no interest at all, 5 - a large amount of interest).  
Nine participants are daily or weekly users of large language models (LLMs, \eg ChatGPT and Claude), and the other three participants are infrequent users. 
}

\begin{table}[]
\caption{Participants involved in the within-subjects study.}
\resizebox{\columnwidth}{!}{%
\begin{tabular}{cccccc}
\hline
ID  & Gender & Age & Year     & Experiences & Freq. of LLMs Usage \\ \hline
P1  & M      & 22  & Graduate & TA \& Home Tutor                & Daily               \\
P2  & M      & 23  & Graduate & TA \& Home Tutor                & Weekly              \\
P3  & M      & 29  & Graduate & N/A                             & Have Tried          \\
P4  & M      & 22  & Graduate & N/A                             & Weekly              \\
P5  & M      & 25  & Graduate & TA \& Home Tutor                & Weekly              \\
P6  & M      & 24  & Graduate & TA                              & Infrequently        \\
P7  & M      & 22  & Undergraduate   & N/A                & Weekly              \\
P8  & M      & 24  & Graduate & TA                              & Daily               \\
P9  & F      & 23  & Graduate & TA                              & Infrequently        \\
P10 & M      & 23  & Graduate & TA                              & Weekly              \\
P11 & F      & 21  & Undergraduate   & TA                              & Daily               \\
P12 & M      & 23  & Graduate & TA                              & Daily               \\ \hline
\end{tabular}%
}
\label{tab:participants}
\end{table}

\subsubsection{Experiment Setup}
\peng{The goal of the within-subjects study is to quantitatively evaluate the \name{}'s effectiveness and user experience compared to a baseline system}. 

\peng{
\textbf{Baseline condition}. 
The baseline system is the ChatGPT web app that uses the GPT-4 model as used in \name{}. 
In the baseline condition, participants can also use the Bing search engine in their web browsers and a markdown editor \footnote{https://github.com/code-farmer-i/vue-markdown-editor} that is identical to the markdown component embedded in \name{}.
}

\peng{
\textbf{\name{} condition}. 
In the \name{} condition, participants are permitted to use Bing and \name{}, but are not allowed to visit the ChatGPT web app or another markdown editor. 
In either condition, participants could freely choose whether, when, and how to use the provided tools. 
}

\peng{
\textbf{Task-system assignment}. 
In the recruitment survey, we ask participants to choose the course topics of Data Structures they are familiar with. We select the two topics that all participants indicate their familiarity with, \ie ``Hash Table'' and ``Introduction and Traversal of Binary Trees''. 
We counter-balance the order of lesson planning tasks and use a system using Latin Square, with three participants in each of the four assignments:  
\begin{itemize}
    \item \name{} - Hash Table, Baseline - Introduction and Traversal of Binary Trees
    \item Baseline - Introduction and Traversal of Binary Trees, \name{} - Hash Table 
    \item \name{} - Introduction and Traversal of Binary Trees, Baseline - Hash Table
    \item Baseline - Hash Table, \name{} - Introduction and Traversal of Binary Trees
\end{itemize}
}


\subsubsection{Tasks and Procedure}
\peng{
Each participants have two lesson-planning tasks. The prompt for each task is: 
\begin{quote}
    The lesson plan is a teacher's detailed description of the course of instruction. 
    Now, you are instructed to act as a teacher who will deliver a lesson about Data Structures - [Topic Name in this task] to second-year university students who are new to this subject. 
    Your class is composed of 70 students whose academic abilities follow a normal distribution. 
    The lesson is scheduled to last for 90 minutes, and you have 40 minutes available to develop your lesson plan. 
\end{quote}
}

\peng{
The whole procedure is illustrated in \autoref{fig:procedure}. Before the day of conducting lesson planning tasks, we sent participants the following materials, which they should spend 15-30 minutes reviewing before the tasks. 
\begin{itemize}
    \item Sections on Hash Tables and Introduction and Traversal of Binary Trees in a digital textbook of Data Structures and Algorithms
    \item A mind map that introduces Gagne's Nine Events.
    \item A document that introduces the basics of Markdown operations. 
    \item Two example lesson plans of course topics different from those in Data Structure and Algorithm. 
\end{itemize}
}

On the day of conducting lesson planning tasks, each three participants that are in the same task-system assigned group come to our lab. 
We briefly introduce Gagne's Nine Events and the considerations for a good lesson plan using the materials sent yesterday.  

\peng{
To motivate them to put forth their best effort in the task, we inform them that an additional 80RMB reward will be given to the best lesson plan among others rated by a seasonal lecturer of Data Structures and Algorithms. 
}

\peng{
In each task, we first introduce the task and demonstrate the assigned system to participants. 
We then help participants to set up the environments of study in our provided computers or their laptops, \ie opening webpages of the baseline system or \name{}, Bing, and digital textbook of the assigned course topic. 
Each participant then independently works on the lesson plan preparation task. 
We allocate 40 minutes for each task and inform them that they can finish earlier or have a few more minutes to complete the task. 
After each task, we ask participants to fill out a questionnaire about their experience in this lesson planning process and perceptions of the used system. Subsequently, we interview them for 5 minutes. 
There is a 10-minute break between two tasks for participants to take a rest. 
Upon completion of two tasks, we conduct a final semi-structured interview with the participants that focuses on their \pzh{preferences on the used systems, perceptions on the generated content, opinions on the features of \name{}, and suggestions for improving it.}
In total, each participant spends around 120 minutes in our experiment and receives 120RMB for compensation. 
}

\subsubsection{Measurements}
We employ a standard 7-point Likert scale (1 - strongly disagree, 7 - strongly agree) to measure the quality of outcome lesson plans, experience in the lesson planning process, and perceptions towards the used system. 

\peng{
\textbf{RQ1: Lesson planning outcome.} 
To assess the quality of the outcomes, we invite a teacher with three years of experience teaching Data Structures and algorithms to score all outcome lesson plans in a shuffle order. 
The aspects of evaluation are adapted from ~\citet{aziz2018implementation}, based on the CIPP model, and the lesson plans are assessed on a 7-point Likert scale to indicate the degree to which each criterion is met. 
For each lesson plan, the teacher assesses it from \pzh{
five key aspects adapted from the CIPP lecture evaluation model \cite{aziz2018implementation} include the alignment of teaching content with learning objectives, facilitation of students' skill acquisition, integration of effective teaching materials, a balance between theoretical and practical activities, and the employment of effective teaching strategies. Sufficient detail has also been added as an aspect to further evaluate the lesson plans.
}
}

\textbf{RQ2: Lesson Planning Process.}  
Drawing from the NASA-TLX survey~\cite{hart1988development}, we pose six questions to measure the workload during the lesson planning process, including the aspects of Mental Demand, Physical Demand, Temporal Demand, Performance, Effort, and Frustration. 
In the interviews, we encourage users to share how the system has increased or decreased their workload.

\textbf{RQ3: Perception of \name{}.} 
First, the ten questions from the \textbf{System Usability Scale} (SUS)~\cite{brooke2013sus} are set in our questionnaire. We divide SUS into and analyze it from three levels: Effectiveness \& Learnability, Use Efficiency, and Satisfaction. 
Then we adapted five questions from Jian's Trust Scale~\cite{jian2000foundations} and investigated how users trust the system by inquiring about the system's vigilance, potential negative impact on teaching, their trust in the system's ethical standards, commitment to users, and the reliability of outputs.
In the interviews, we encourage them to share their experiences with and preferences for each tool (\eg search engine; ChatGPT web app; \name{}) and each component (\eg different pre-set activities and the editor in \name{}).
We further inquire about their reasons for trusting or distrusting the system.

\subsection{Expert interviews}
Following the within-subjects study, to gather additional feedback to enhance the generalization of our findings, we conducted think-aloud studies and semi-structured interviews with six teachers.

\subsubsection{Participants}
The six teachers (female=2, male=4) include five novice teachers with teaching experience ranging from 1 to 3 years ($M=2$), and one experienced teacher with 15 years of teaching experience. Participants E1-E3, who have been involved in the formative study, teach at the university level. E4 and E5 are primary school teachers, and E6 teaches at a senior high school. Detailed information about their teaching subjects in the current year (the same as what topics they choose to test on our system) are presented in~\autoref{tab:experts}.
\begin{table*}[]
\caption{Experts involved in user study.}
\label{tab:experts}
\begin{tabular}{ccccc}
\hline
ID & Gender & Educational Career length & Educational Level  & Subject(s)                              \\ \hline
E1 & F      & 15 years           & Senior High School & English                                 \\
E2 & M      & 3 years            & Elementary School  & Mathematics                             \\
E3 & M      & 1 year             & Elementary School  & Science                                 \\
E4 & F      & 3 years            & University         & Korean Intensive Reading                \\
E5 & M      & 2 years            & University         & Natural Language Processing             \\
E6 & M      & 2 years            & University         & Convex Optimization \& Computer Network \\ \hline
\end{tabular}%

\end{table*}
\subsubsection{Method}

We conduct a 60-minute interview for every expert. Initially, we introduce the participants to the background and goal of our research. For E4-E6, we present Gagne's Nine Events of Instruction and confirm their full understanding of the concept. Subsequently, we give them a 10-minute tutorial on how to use \name{}. They were given 30 minutes to complete the following tasks, being asked to think aloud. \fhxr{The examples of expert's interactions with \name{} is provided in supplementary material 3.}
\begin{itemize}
    \item \textbf{Small Task 1.} Generate lesson goals and the outline and set up instructional events.
    \item \textbf{Small Task 2.} Refine a section. Add an example of a question generated by LLMs.
    \item \textbf{Exploration.} Explore and refine any areas that they are interested in.
\end{itemize}

Finally, we conduct a 15-minute semi-structured interview, the fixed questions are shown below.

\textbf{RQ1: Lesson planning outcome.}
\begin{itemize}
    \item Please rate the quality of lesson plans created by \name{}, compared to those you prepare during your usual planning routine.
    \item Can \name{} help you manage the teaching process effectively?
\end{itemize}

\textbf{RQ2: Lesson Planning Process.}
\begin{itemize}
    \item Does our system impose additional cognitive load on you, such as memory or thought burden, compared to preparing lesson plans without it?
\end{itemize}

\textbf{RQ3: Perception of \name{}.}
\begin{itemize}
    \item Is \name{} easy to handle? Can you use \name{} smoothly without any technical support?
    \item Which feature (including goal-setting, outline generating, pre-set activities, and output refinement) provided by \name{} help you most?
    \item Please provide some suggestions for \name{}.
\end{itemize}

\section{Analyses and Results}
\peng{
\pzh{In this section, we present the quantitative and qualitative results for each research question (RQ).} 
As for the items measured on a 7-point Likert scale about the lesson planning outcome (RQ1), lesson planning process (RQ2), and perceptions with the system (RQ3), we employ the Wilcoxon signed-rank tests~\cite{woolson2007wilcoxon} compare the differences between the \name{} and baseline condition in the within-subjects study. 
For the qualitative data in the within-subjects study and expert interviews, two authors transcribe the audio into text scripts and conduct a thematic analysis on these scripts.
they first familiarize themselves by reviewing all the text scripts independently. 
After several rounds of coding with comparison and discussion, they finalize the codes of all the interview data. 
We count the occurrences of codes and incorporate these qualitative findings in the following presentation of our results.
}
\peng{
\subsection{Lesson Planning Outcome (RQ1)}
}

\begin{figure*}[tbp]
  \centering
  \includegraphics[width=0.9\textwidth]{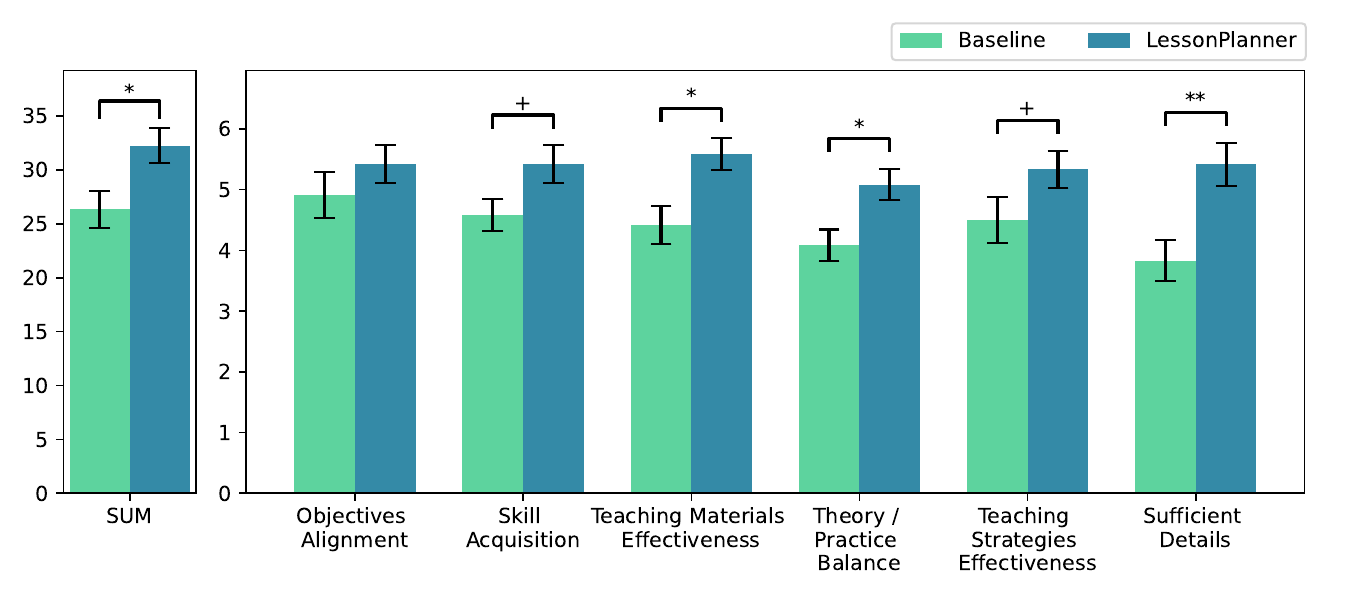}
  \caption{RQ1 results regarding the lesson plan outcome evaluated by experienced teachers in six different aspects and total scores. ***: p< 0.001, **: p< 0.01, *: p< 0.05, +: p<0.1.}
  \Description{RQ1 results regarding the lesson plan outcome evaluated by experienced teachers in six different aspects and total scores. ***: p< 0.001, **: p< 0.01, *: p< 0.05, +: p<0.1.}
  \label{fig:RQ1}
\end{figure*}
\peng{
\autoref{fig:RQ1} shows the quality of outcome lesson plans from the within-subjects study\fhxr{, and the the lesson plan designed by P5 is provided in supplementary material 2}. 
Overall, the lesson plans created using \name{} (\pzh{$\textbf{M}ean = 32.25, SD = 5.43$}) are rated significantly better than those developed with the baseline system (\pzh{$M = 26.33, SD = 5.588$}) in terms of the total scores; $W=10.5, p=0.02$. 
Regarding each aspect of the outcome lesson plans, 
\name{} (\pzh{$M = 5.42, SD = 1.037$}) does not show a significant advantage over the baseline system (\pzh{$M = 4.92, SD = 1.256$}) regarding aligning the teaching content to the objectives; $W=15.5, p=0.40$. 
However, 
compared with the baseline condition, the lesson plans from the \name{} conditions have a tendency to perform better in facilitating students' skill acquisition (\pzh{\name{}: $M = 5.42, SD = 1.04$; baseline: $M = 4.58, SD = 0.86$}; $W=7, p=0.058$) and incorporating effective teaching materials (\pzh{\name{}: $M = 5.33, SD = 1.03$; baseline: $M = 4.50, SD = 1.26$}; $W=13.50, p=0.070$). 
Lesson plans constructed with \name{} are significantly better in terms of balancing the theoretical and practical content (\pzh{\name{}: $M = 5.08, SD = 0.86$; baseline: $M = 4.08, SD = 0.86$}; $W=8, p=0.039$), using effective teaching strategies (\pzh{\name{}:$M = 5.58, SD = 0.86$; baseline: $M = 4.42, SD = 1.04$ }; $W=7, p=0.034$), and providing sufficient \textbf{details} on the teaching materials (\pzh{\name{}: $M = 5.42, SD = 1.19$; baseline: $M = 3.83, SD = 1.14$}; $W=3, p=0.007$). 
Our participants in the within-subjects study and teachers in the expert interview share how \name{} affect their lesson planning outcome, which is summarized below. 
}

\peng{
\subsubsection{The Impact of Structured Outlines on Overall Lesson Plan Quality}
Four participants (P1, P2, P6, P12) mentioned that the structured outlines helped improve the quality of their lesson plans. 
As the generated outlines take into account various lesson goals and Gagne's Nine Events, \textit{``\name{} made my lesson plan more standardized''} (P6), and \textit{"it has a more reasonable instructional structure, compared to most resources I found on the Internet"} (P12). 
Furthermore, the block-based structure strengthens users' grasp of the overall lesson plans. 
\textit{``It (\name{}) makes me more clear about what should I do in the different stages of this lesson''} (E5). 
However, E3 offered a different perspective on the impact of the structured outline.
\textit{``When I prepare lessons, I already have a clear framework of this lesson (elementary science) in mind. It is synthesized from many lesson plans I've read. Compared to my own framework, I personally find the generated one less effective''}(E3). 
It reveals that the generated lesson plan outline in \name{}, which is based on Gagne's Nine Events, may be less effective in specialized subject domains. 
}
\subsubsection{Experts' Opinions on the Impact of Event Hints on the Teaching Content}
The teachers in the expert interviews confirmed that the \pzh{tags that give} hints of Gagne's Nine Events in \name{} can guide them to prepare a lesson plan with more effective teaching content. 
E6 mentioned, \textit{``without the system, I'd always forget to put some important content in the lesson plans. The hints act like reminders''}. 
E1 also affirmed this point, 
\textit{``(Without \name{},) I wouldn't keep so many tags (instructional events)in my mind. These constantly visible tags help me think and explore the possibilities of conducting my class from various perspectives''}.
However, not all experts think the Events Hints generated in the outlines are useful. 
E2 and E3 complained that some hints were useless for them. 
E2 said, \textit{``Elementary students have limited comprehension, and a single math lesson is impossible to encompass all nine events. I have to adjust the events in each part to fit my need''}.

\peng{
In all, we find that \name{} significantly improves participants' quality of lesson plans in the within-subjects study. 
Participants and teachers generally value the structured lesson plan outline and hints of Gagne's Nine Events. 
Nevertheless, these events may not be always necessary if teachers have their own lesson plan structures and if they want to focus on specific events in a lesson.
}

\peng{
\subsection{Lesson Planning Process (RQ2)}
}

\begin{figure}[tbp]
  \centering
  \includegraphics[width=0.49\textwidth]{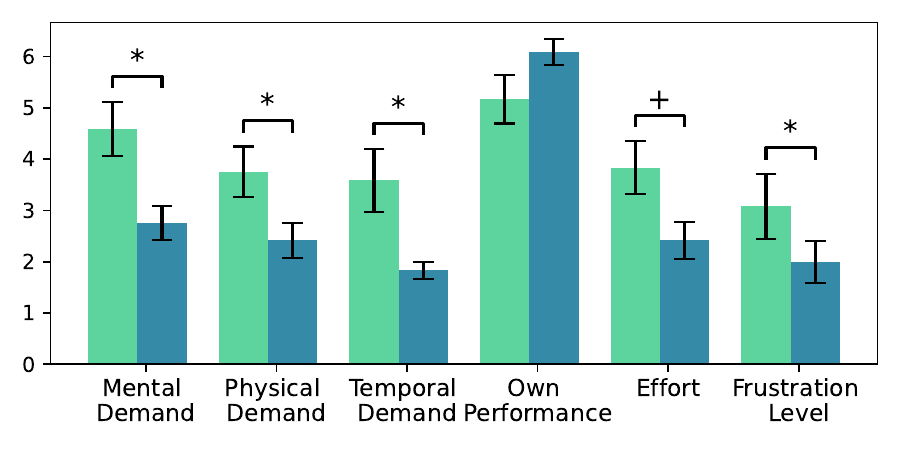}
  \caption{RQ2 results regarding the test of task load. ***: p< 0.001, **: p< 0.01, *: p< 0.05, +: p<0.1.}
  \label{fig:RQ2}
  \Description{RQ2 results regarding the test of task load. ***: p< 0.001, **: p< 0.01, *: p< 0.05, +: p<0.1.}
\end{figure}

\peng{
\autoref{fig:RQ2} shows the statistical results of the participants' perceived task workload in the lesson planning process with \name{} and the baseline system. 
On average, participants perceived significantly less mental demand (\pzh{\name{}: $M = 2.75, SD = 1.09$; baseline: $M = 4.58, SD = 1.75; W=7.00, p=0.035$}), physical demand (\pzh{\name{}: $M = 2.42, SD = 1.11$; baseline: $M = 3.75, SD = 1.64; W=2, p=0.024$}), temporal demand (\pzh{\name{}: $M = 1.83, SD = 0.55$; baseline: $M = 3.58, SD = 2.02; W=1, p=0.027$}), and frustration (\pzh{\name{}: $M = 2, SD = 1.35$; baseline: $M = 3.08, SD = 2.10; W=6, p=0.046$}) when they prepare lesson plans with \name{} than with the baseline system. 
Besides, participants tend to perceive less effort spent in the lesson planning process when they are with \name{} (\pzh{$M=2.42, SD=1.19$}) than with the baseline system (\pzh{$M=3.83, SD=1.72$}); $W=12.5, p=0.066$. 
There is no significant difference regarding their perceived task performance in lesson planning between the \name{} (\pzh{$M=6.08, SD=0.86$}) and baseline conditions (\pzh{$M=5.17, SD=1.57$}); $W=14.5, p=0.175$. 
Nine participants believe that utilizing our system requires less workload compared to the baseline in the interview after two tasks. 
\pzh{For example, P12 said \textit{``Using ChatGPT can be quite skill-intensive, and often the first response you receive may not align with your expectations. \name{}'s first response is of high quality, which significantly eases the mental and physical demand on me.''}}. 
} 


\peng{
\subsubsection{Experts' Opinions on the Workload of Preparing Lesson Plans with \name{}}
Our within-subjects user study showcases that \name{} can reduce task workload compared to the baseline interface. In the expert interviews, we are more concerned about what brings workload to teachers when planning lessons with \name{}. 
We found that the workload mainly arises from the selection, refinement, and verification of LLMs' outputs. 
First, E4 and E6 indicated that the process of selecting output imposed an additional burden. 
Interestingly, they both reported that this burden was beneficial, as reported by E6,  \textit{"The system offers varied content, allowing me to select anything I am satisfied with. This process naturally demands more effort. Before, I could only stick to one basic idea"}. 
Second, E3 and E4 noted that they invested the majority of their time in ensuring that the system accurately produced a small segment of the content, such as formulas, a suitable example, and so on. E3 complained that \textit{"When I realized that the results provided by this generative model were not what I was looking for, I kept engaging in conversing with it and iterating the generated results. Clearly, it takes up more time"}.
Third, the experts emphasized that checking the accuracy of the content (E4, E5) and making sure it was appropriate for the student's current learning stage (E2, E3) also contributed to their workload. 
For example, \pzh{E2 stated, \textit{``"In this example, it (\name{}) uses the area calculation formula for triangles as an introduction, asking students to think about the formula for the area of parallelograms. Yet, the students don't know how to calculate the area of triangles because it is the content in the next chapter... Having to check every output makes me drained.''}}. 
}
\peng{
\subsection{Perception of \name{} (RQ3)}
}
\begin{figure*}[tbp]
  \centering
  \includegraphics[width=0.9\textwidth]{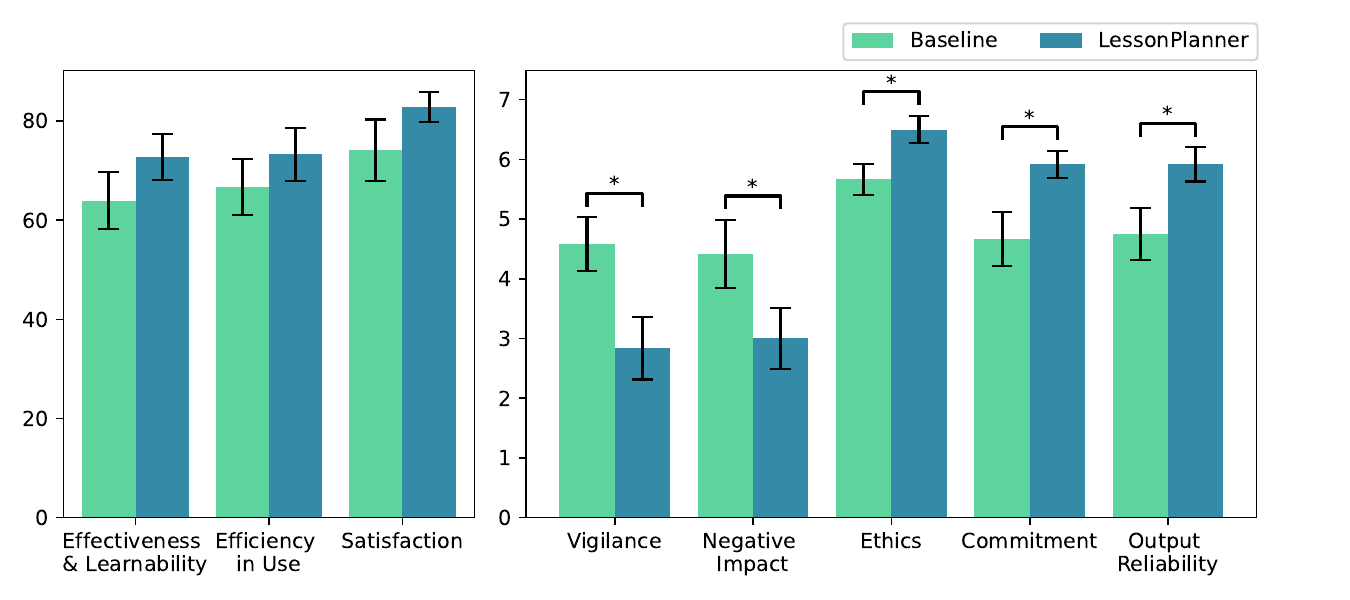}
  \caption{RQ3 results regarding the test of system usability and participants' trust in \name{}. ***: p< 0.001, **: p< 0.01, *: p< 0.05, +: p<0.1.}
  \Description{RQ3 results regarding the test of system usability and participants' trust in \name{}. ***: p< 0.001, **: p< 0.01, *: p< 0.05, +: p<0.1.}
  \label{fig:RQ3}
\end{figure*}

\peng{
\autoref{fig:RQ3} shows the participants' ratings on the usability of and their trust on \name{} and the baseline system in the within-subjects user study. 
Overall, participants gave a higher score for \name{} than that for the baseline system regarding the system's Effectiveness \& Learnability (\pzh{\name{}: $M = 72.70, SD = 15.97$; baseline: $M = 63.90, SD = 19.90; W=21, p=0.286$}), Efficiency in Use (\pzh{\name{}: $M = 73.30, SD = 18.50$; baseline: $M = 66.70, SD = 19.78; W=27, p=0.380$}), and Satisfaction (\pzh{\name{}: $M = 82.90, SD = 10.45$; baseline: $M = 74.10, SD = 21.62; W=20, p=0.246$}), but the differences are not significant.  
As for the trust in the system, compared to the baseline system, participants generally have significantly fewer concerns regarding the \name{}'s vigilance (\pzh{\name{}: $M = 2.83, SD = 1.72$; baseline: $M = 4.58, SD = 1.50; W=4, p=0.028$}) and potential negative impact on teaching (\pzh{\name{}: $M = 3.00, SD = 1.68$; baseline: $M = 4.42, SD = 1.90; W=4, p=0.048$}). 
Participants also have a significantly higher level of trust on \name{}'s ethical standards (\pzh{\name{}: $M = 6.50, SD = 0.76$; baseline: $M = 5.67, SD = 0.85; W=3.5, p=0.020$}), commitment to user (\pzh{\name{}: $M = 5.92, SD = 0.76$; baseline: $M = 4.67, SD = 1.49; W=3.5, p=0.012$}), and reliability of outputs (\pzh{\name{}: $M = 5.91, SD = 0.95$; baseline: $M = 4.75, SD = 1.42; W=2, p=0.024$}). 
Participants and teachers actively comment on the usability of \name{} and their trust or concerns about its generated content in the interviews, which we summarize below. 
}




\peng{
\subsubsection{Experts' Gained Insights from \name{}}
Our teachers in the expert interviews appreciated that \name{} can provide inspiring content for constructing lesson plans.
Specifically, all teachers agreed that the "Generate an Example in Detailed" feature is well-designed and useful, as it provides them with new perspectives for lesson design. 
For instance, E2 mentioned, \textit{``Sometimes I come up with an example but I am uncertain if it's appropriate. I am also often not sure from which angle I should present the example to the students. \pzh{I can get other examples from the system as a reference to my example and get guidance on how to present it}''}. 
Moreover, when using "Construct multiple choice or fill-in-the-blank questions" or "Construct group discussion topics" in their small tasks of lesson planning, four experts expressed that the system output was "very inspiring". 
\pzh{For example, E6 stated, "\textit{I had never considered incorporating discussion questions into the (Computer Networks) course, but now it has generated a great topic I think. Discussing this topic is beneficial for students, actually.}"}.
These qualitative findings once again confirm that LLMs are competent tools for promoting users' creativity~\cite{davis2022investigating}. Novice teachers greatly benefit from the insights generated by LLMs and use them to broaden their thinking.
}

\peng{\subsubsection{Diverse Attitudes to the Reliability of \name{}}
Eight participants in the user study indicated that they were more confident with the reliability of the generated content in \name{} compared to that in the baseline. 
For example, P9 and P12 accounted for their increased confidence in the preset prompts and structured output, which make \name{} seem professional and reliable. 
In contrast, one participant, P3, doubted the preset prompts, \textit{``I would only trust the system if I can access all the details of the prompts.''}. 
}

\peng{
Teachers in the expert interviews have different opinions on the reliability of the LLM's output.
For example, the most frequently debated issue was the credibility of the ``Generate definition'' activity. 
E5, who teaches Computer Network and Convex Optimization at the university level, believed that the generated definitions were accurate and could be directly utilized in his lessons. 
Conversely, 
E2 and E3, who are instructors in elementary school, contended that the definitions did not correspond with the student's current stage of learning, making them nearly impracticable for the classroom. 
\pzh{This conflicting view could be due to that our \name{} was primarily designed for supporting teachers in university in our study, but it reveals the need to tailor the generated content to the knowledge level of targeted students, which we will discuss in the Discussion section.}
}

\peng{
Another interesting finding is that users' trust in \name{} could be affected by the level of their familiarity with LLMs. 
Participants who interact with LLMs daily or weekly ($M=78.0, SD=17.20$) generally show a higher degree of trust in the system than those who use it less frequently ($M=68.3, SD=14.55$).
Additionally, in the expert interviews, teachers (E2, E4) who do not know how LLMs work and rarely used LLMs before, expressed less satisfaction with \name{}, which could be due to their \textit{inappropriate trust}~\cite{qian2024take}. 
For instance, E4 stated, \textit{"The system needs to give clear sources for its generated examples, such as from newspaper. I would not count on an example without credible sources, because it might be misleading for students.
"}
}

\peng{
\subsubsection{Room for Improving \name{}'s Usability}
Eight participants in the user study suggested that our system had room for improved usability. 
First, the embedded editor was considered "not so intelligent" by four participants (P2, P4, P8, and P10).
For example, P2 mentioned, \textit{``I usually use shortcuts, but here, I have to use the mouse to click. It is inconvenient.''} 
Second, \name{} interaction design may not match the routine way to use LLM. 
P1 and P9, who are familiar with the ChatGPT web app, struggled with our system's conversation design. 
\textit{``I am used to scrolling up to see what it (LLM) responded to before, but it does not work in this system. I miss that the way I do all the time with ChatGPT''} (P1).
}

\peng{
In the expert interviews, we further inquired about the learnability of \name{}.
All teachers believed that the system was easy to use, but they reminded that it should have a clear tutorial to walk through the system before using it. 
Without a tutorial, they might be confused about the system's functions. 
For example, E2 and E5 claimed that they were confused by the labels on buttons without detailed descriptions of their functions, input, and output. 
E5 suggested that \textit{``when a user enters this system for the first time, providing a new user tutorial or a product tour could be helpful, as many websites do}''.
}
\section{Discussion}
\peng{
In this work, we design and develop \name{}, an interactive system that facilitates novice teachers to prepare lesson plans with large language models (LLMs). 
Our within-subjects study with 12 participants compared to a ChatGPT interface and our expert interviews with six teachers demonstrate the usefulness of \name{} in improving the outcome and reducing the task load of preparing lesson plans. 
Our user study findings also highlight the imitations, and opportunities of \name{}, which we summarize as design considerations for future interactive systems that support teachers with LLMs. 
}
\subsection{Design Considerations}
\peng{
\subsubsection{Base the generated content on reliable knowledge databases}
Both participants and experts acknowledged the remarkable generative capabilities of LLMs for supporting lesson-planning tasks. 
\name{} can assist teachers in producing various types of insightful and creative teaching materials and suggesting strategies to deliver these materials.
Nevertheless, the reliability of the LLM used in \name{} was still questioned by some participants and experts (E2-E5), primarily because of the hallucination of LLMs~\cite{zhang2023siren} and the limitation of not being able to automatically access web resources. 
Furthermore, P12 and E3 expressed a continued preference for utilizing search engines to search for materials, though LLMs are equipped in \name{}.
\fhxr{On the one hand, the generated content may not match different educational levels, for example, including knowledge that elementary students have not yet grasped in previous lessons (E2). This problem stems from teachers providing only a brief description of prior knowledge in a single line on the Goal-setting page (\autoref{fig:goal}), which is insufficient. 
\zhenhui{To improve the alignment of the generated content with the students' educational levels, future work could construct a detailed knowledge graph of courses at different levels and recommend prior knowledge of current course topics to guide the \name{}.} 
On the other hand,}
the preference stemmed from their belief that despite the unstructured results returned from search engines, the online materials published by other teachers are more professional, compared to those generated by LLMs. 
Hence, content generated by LLMs should not constitute the entire lesson plan but rather serve as a supplementary resource to complement the lesson plan. 
This principle is fundamental in our motivation, design, and implementation of \name{}.
To maximize the potential of traditional resources, web resources, and generated content within the system, we suggest that future lesson planning systems should integrate reliable knowledge databases, either prepared by teachers or curated from web resources, into the prompts to LLMs. 
There are already many ways that help us do this, such as LangChain~\footnote{\url{https://github.com/langchain-ai/langchain}}, Microsoft Copilot, and the recent advances in Retrieval-Augmented Generation techniques\pzh{~\cite{lewis2020retrieval, cai2022recent}}. 
With the assistance of external resources, teachers can enhance their confidence in the system and decrease the workload they spend on verifying the generated content.
}

\peng{
\subsubsection{Allow flexible customization of prompts to LLMs.}
When designing and developing \name{}, we spend a lot of effort in providing flexible user control (DG4) to the generated content, \eg users are able to alter or cooperate with LLMs to refine the lesson goals in the Goal-Setting page, and users can freely modify all generated content of the lesson plan in the Lesson-Planning page. 
However, users hold elevated expectations regarding the customization of the interactive features with LLMs. 
For example, participants expressed a desire to see and adjust our built-in prompts of the default buttons, which would make the generated content align more closely with their expectations.
Teachers in the expert interviews further highlighted that Instructional Events used in the LLM prompt for initializing the lesson plan may need to be adjustable by users as well. 
Potential solutions to address these expectations are incorporating additional predefined teaching scenarios (\pzh{\eg a lesson after midterm exams}) offered by developers and enabling users to customize the system's predefined prompts. 
Another approach is to automate the prompting process~\cite{liu2023pre} by utilizing course meta-data to provide different predefined functionalities tailored to users.
}

\peng{
\subsubsection{Integrate multi-modal generated content}
The findings of our study demonstrate that planning with \name{} leads to a notable enhancement in the quality of lesson plans and a reduction in teachers' workload, which can be attributed to the abundant textual content produced and structured by the LLM. 
An observation identified in the expert interviews is that \name{} demonstrates its interest in \pzh{using multi-modal materials to teach particular knowledge points.}
For instance, it attempts to draw a binary tree through a combination of text and symbols or directs the teacher to present a video on specific topics or display images in the classroom.
The aforementioned phenomenon reminds us that integrating other multi-modal models, such as text-to-image generation models~\cite{zhang2023text} and text-to-video generation models~\cite{cho2024sora}, with LLMs might make the content of lesson plans richer and save time by reducing the need to gather supplementary materials from the web.
These two points were further confirmed by the experts in the interviews, where they hoped to insert generated images into the lesson plans. 
}

\peng{
\subsection{Generality}
In the within-subjects study, \name{} demonstrated strong performance, as no participants raised concerns about the suitability of the content delivered by LLM for students at their current stage.
The teachers \zhenhui{of different educational levels} in the expert interviews have expressed overall satisfaction with the system's performance, despite that 
\zhenhui{the examples in the prompt templates of \name{} are about the subjects of computer science for university students. }
This suggests that our system has the potential to be adapted to various educational levels and subjects.
\fhxr{
\zhenhui{However, our teachers reported that they occasionally got generated definitions or examples unaligned with the students' knowledge background.} 
To mitigate this issue, two feasible solutions can be considered.
First, the simplest and most practical approach \zhenhui{could be encouraging} users to customize examples or definitions as templates and embed them into the original prompts. 
This will guide LLMs to generate content that is aligned with the provided templates.
Second, the Chain-of-Thought (CoT)~\cite{wei2022chain} technique could be employed in prompt engineering, instead of providing \zhenhui{fixed} examples (\eg data structures) which might lead to content that is inappropriate for the educational level or subject. 
For example, the prompt shown in~\autoref{example_prompt_with_input_and_output} could be edited as below.}
\begin{itemize}
    \item \fhxr{
    Q1: I will instruct the course of \{lesson name\} for students in \{students' \zhenhui{educational level}\}. Please provide the names of three key concepts that students may need to learn for this course.}
    \item \fhxr{
    Q2 (after receiving the response to Q1 from the LLMs): Provide the specific definition of the first concept that is suitable for the \{students' \zhenhui{educational level}\}. The response must conform to the following format.\\}
    \fhxr{Input:\\}
    \fhxr{Context: the name of the concept\\}
    \fhxr{Output:**the name of the concept **}
    \fhxr{-Definition: The definition of the concept.}
\end{itemize}
\fhxr{
Moreover, \name{} serves not only as an assistant in preparing lesson plans for courses but also as a platform helping novice teachers promote their critical thinking and delve deeper into the subjects. It assists educators in gaining the knowledge of developing a lesson based on educational theory in a systematic way, thus improving their pedagogical abilities.
}
}

\peng{
\subsection{Limitations and Future Work}
Our work has several limitations that call for future work. 
First, all participants in our within-subjects study majored in subjects related to computer science, which helps to control the topics of planned lessons in the study. 
\zhenhui{While we qualitatively evaluate \name{} with six experts teaching different subjects at various educational levels, we lack quantitative findings on \name{}'s effectiveness in facilitating users with diverse backgrounds in planning lessons on other subjects. }
Second, \name{}, which includes detailed instructional events and delivery methods, is presently oriented towards assisting novice teachers. 
In the future, we will evaluate and customize \name{} with teachers with diverse backgrounds and teaching experience. 
Third, while the output markdown file of the lesson plans from \name{} can serve as the intermediate materials to deliver a course, teachers usually need to prepare other digital materials (\eg slides) upon the lesson plans in their teaching practices. 
Future researchers can explore the application of deep learning techniques to generate slides~\cite{zheng2022telling, fu2022doc2ppt} based on the output lesson plans from \name{} or support users to prepare a lesson in Powerpoint or Google slides embedded with LLMs. 
Fourth, we invite a teacher who is experienced with the course topics to evaluate the outcome of lesson plans in the within-subjects study.
However, a more ideal way to assess the effectiveness of a lesson plan would be to enact the plan in a real-world lesson and collect feedback from students 
\fhxr{and teachers.} 
In the future, we plan to invite teachers to use \name{} to prepare for the lessons they are going to teach and evaluate \name{}'s effectiveness after the outcome lesson plans are enacted in a course. 
\fhxr{Recordings of the lessons and post-lesson interviews with the teachers will also be collected and analyzed to ensure \name{}'s practical usability and benefits for educators.}
\zhenhui{Last but not least, to generate an initial lesson plan, we included all nine events in the prompt. While nine events are generally helpful, not all events are applicable to one specific lesson. Future iteration of \name{} should allow the users to specify the events they would like to incorporate into their course before prompting the LLMs to generate a lesson plan.}
}

\section{Conclusion}
\peng{
In this paper, we designed and developed an interactive system, \name{}, to support novice teachers in interactively constructing lesson plans using LLM-generated content grounded on Gagne's nine events. 
\name{} is capable of initializing lesson goals and outlines, generating activities and materials that adapt to planned teaching events, and offering flexible user control to customize the lesson plan and query the LLM. 
The within-subjects experiment with 12 participants demonstrates that \name{} leads to an enhancement in the quality of the lesson plan outcome and eases their workload during the preparation process. We further conducted expert interviews with six teachers who highlighted the potential of utilizing it to support novice teachers at various educational levels and in diverse subjects. 
We discuss design considerations and insights derived from the user study for leveraging generative models to support lesson-planning tasks.
}

\fhxr{\section{acknowledgement}
This work is supported by the Young Scientists Fund of the Na- tional Natural Science Foundation of China (NSFC) with Grant No.: 62202509, NSFC Grant No.: U22B2060, and the General Projects Fund of the Natural Science Foundation of Guangdong Province in China with Grant No. 2024A1515012226.}

\bibliographystyle{ACM-Reference-Format}
\bibliography{references}

\newpage

\appendix

\section{Interfaces of \name{} High-Fidelity Prototype in Formative Study}

\begin{figure}[!b]
\centering
\includegraphics[width=0.49\textwidth]{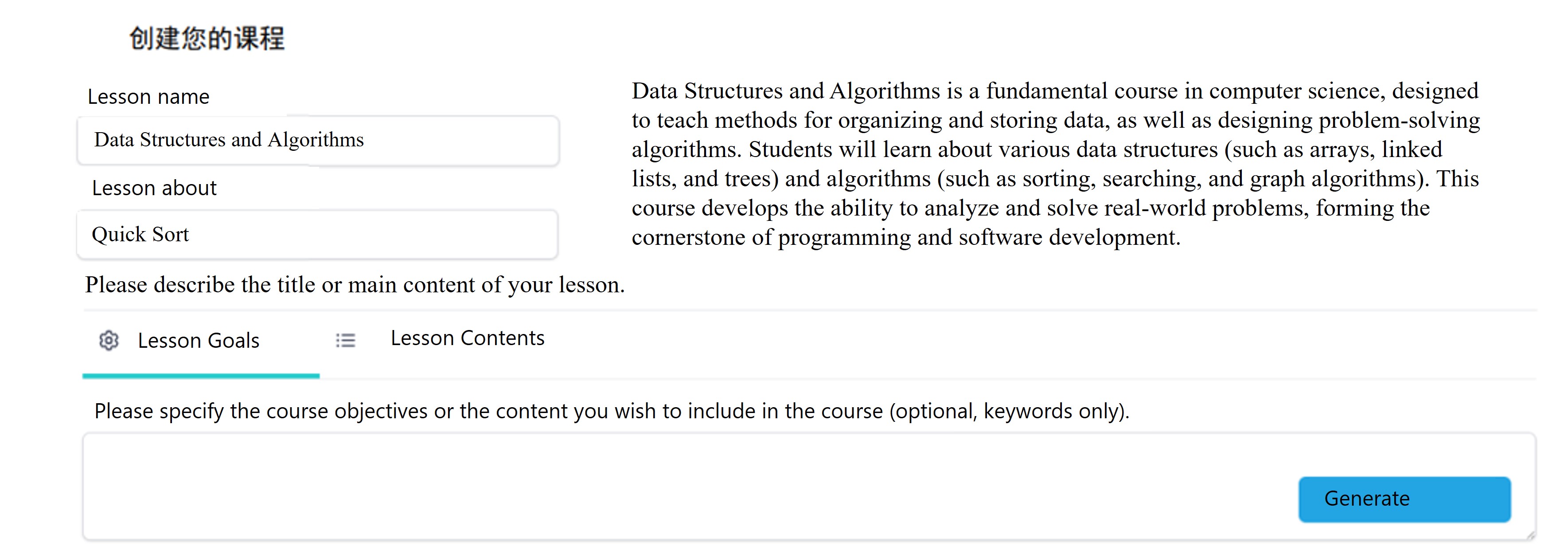}
\caption{\fhxr{The page of meta-data collection.}}
\Description{The page of meta-data collection.}
\label{fig:a1}
\end{figure}

\begin{figure}[]
\centering
\includegraphics[width=0.49\textwidth]{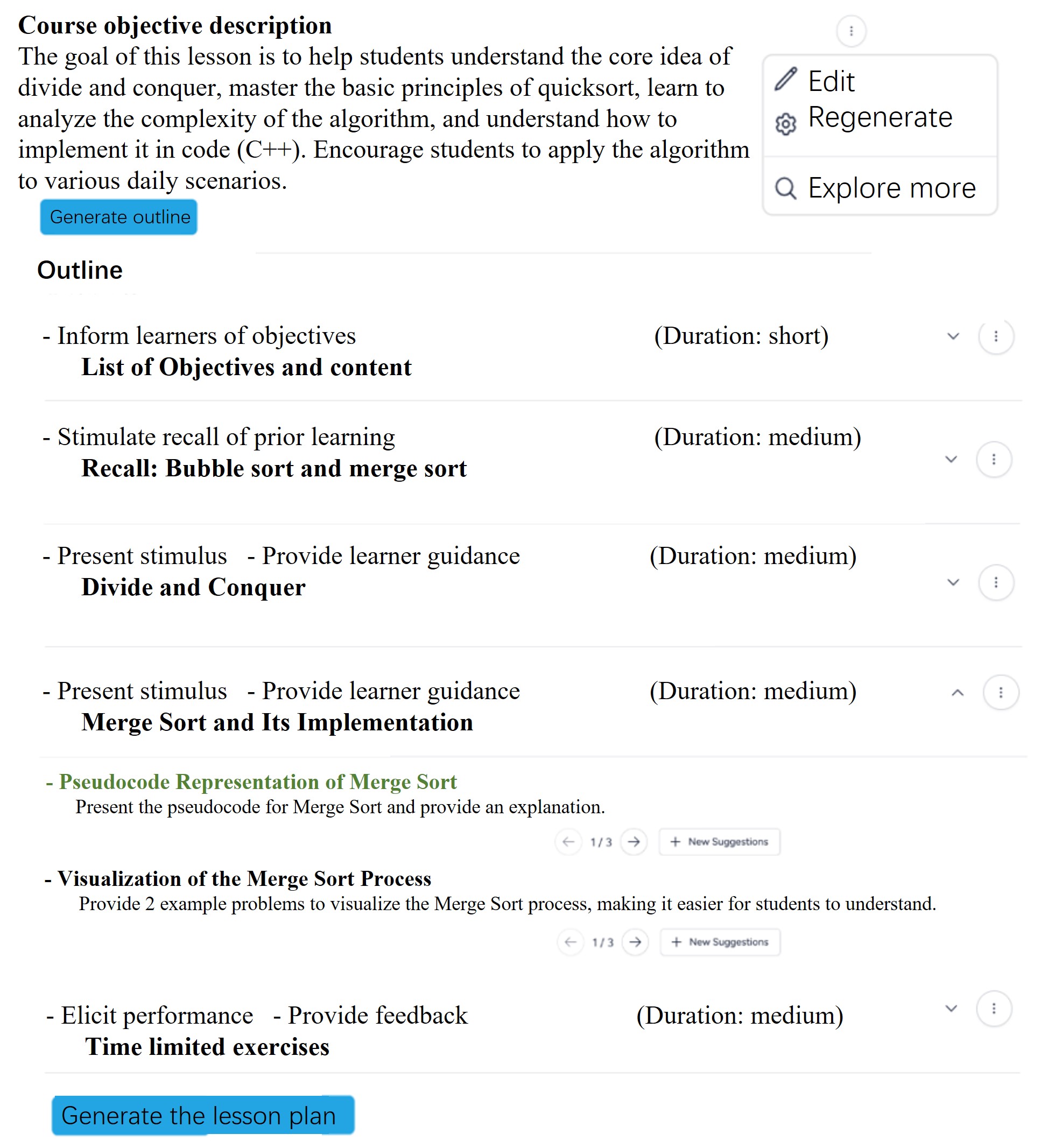}
\caption{\fhxr{The page of outline overview.}}
\Description{The page of outline overview.}
\label{fig:a2}
\end{figure}

\begin{figure*}[]
  \centering
  \includegraphics[width=0.5\textwidth]{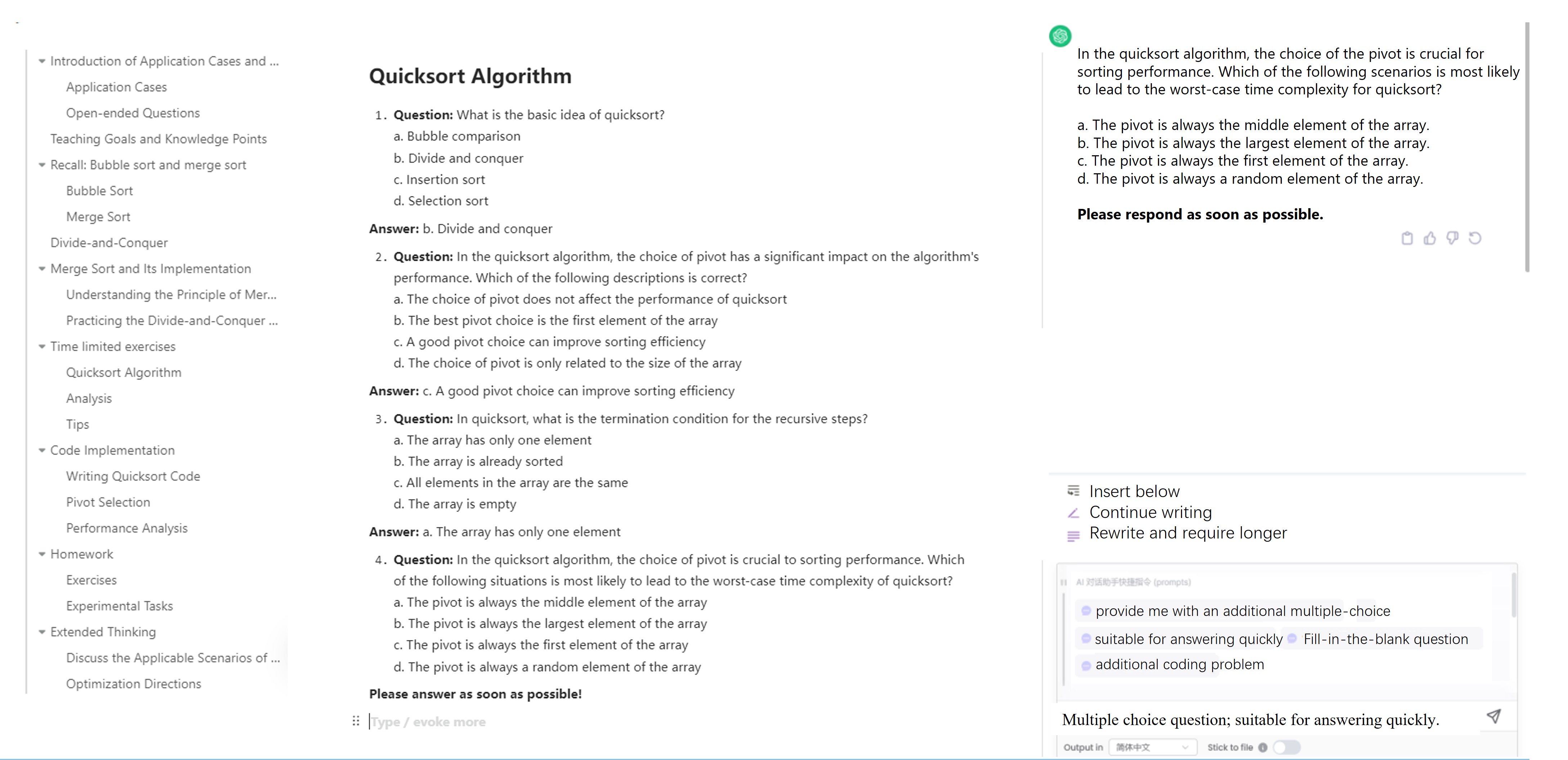}
  \caption{\fhxr{The editing page with LLM sidebar.}}
  \Description{The editing page with LLM sidebar.}
  \label{fig:a3}
\end{figure*}

\begin{figure*}[]
  \centering
  \includegraphics[width=1\textwidth]{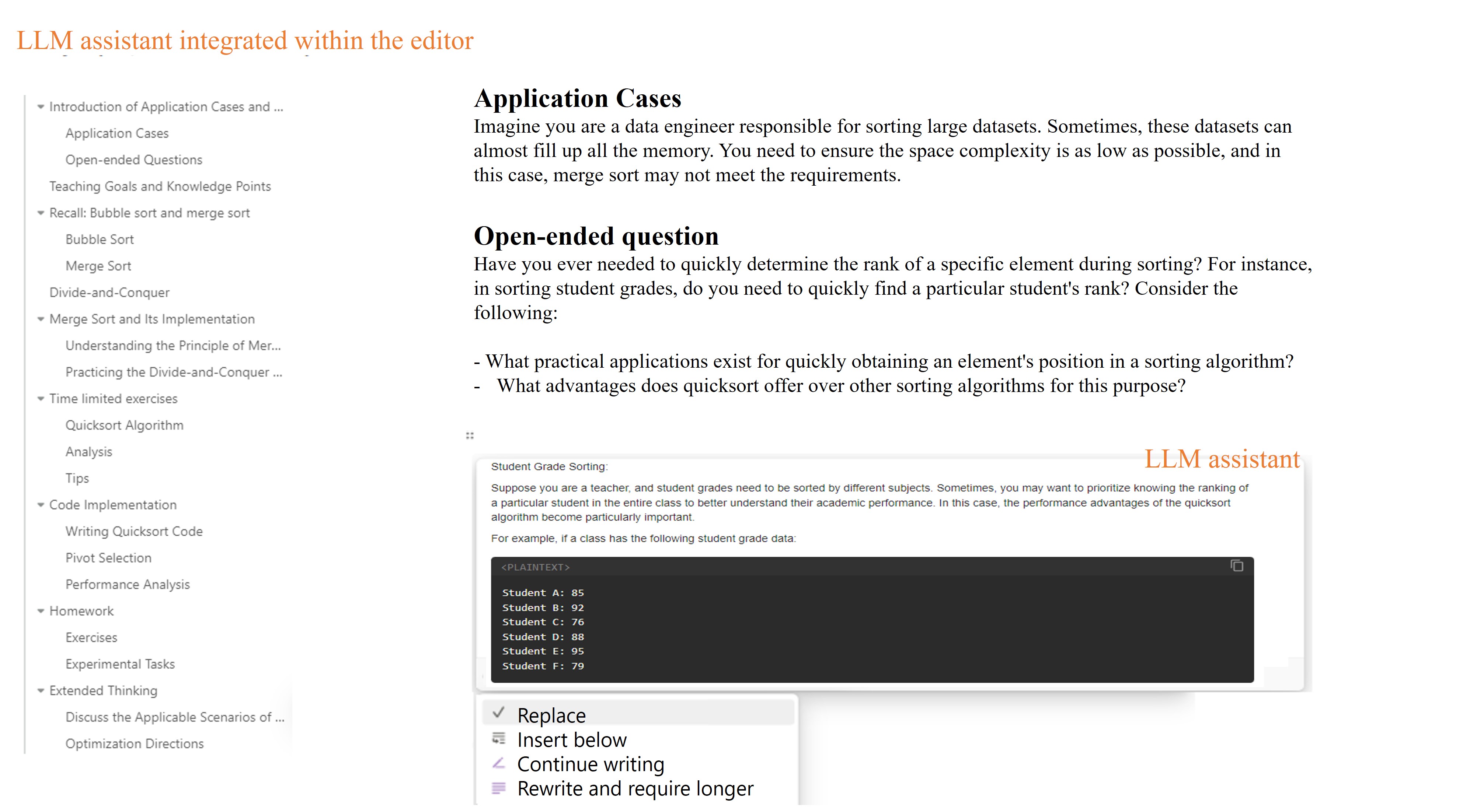}
  \caption{\fhxr{The editing page with embedded LLM assistant.}}
  \Description{The editing page with embedded LLM assistant.}
  \label{fig:a4}
\end{figure*}

The original version used for the actual formative study is in Chinese. We translate all the text in the interface into English completely, as shown in~\autoref{fig:a1}-~\autoref{fig:a4}.

\end{document}